\documentclass[usenatbib]{mnras}
\usepackage[english]{babel}

\usepackage{amsfonts}
\usepackage{amsmath}
\usepackage{amssymb}
\usepackage{xcolor}
\usepackage{enumitem}
\usepackage{graphicx}
\usepackage{gensymb}
\usepackage{fancyhdr}
\usepackage{times}
\usepackage{widetext}
\usepackage{ulem}
\usepackage{caption}
\usepackage{subfigure}
\usepackage{url}

\graphicspath{{./Figures/}}

\begin{document}

\title[Deep extragalactic surveys with JWST]{Maximising the power of deep extragalactic imaging surveys with the James Webb Space Telescope}

\author[Kemp et al.]
{T.\,W. Kemp$^1$\thanks{tkemp@roe.ac.uk}, J.\,S. Dunlop$^1$, R.\,J. McLure$^1$, C. Schreiber$^2$, A.\,C. Carnall$^1$, F. Cullen$^1$\\
  $^1$Institute for Astronomy, University of Edinburgh, Blackford Hill, Edinburgh, EH9 3HJ, UK\\
$^2$Department of Physics, University of Oxford, Keble Road, Oxford OX1 3RH, UK}

\maketitle
\begin{abstract}
We present a new analysis of the potential power of deep, near-infrared, imaging surveys with the {\it James Webb Space Telescope} ({\it JWST})
to improve our
knowledge of galaxy evolution. In this work we properly
simulate what can be achieved with {\it realistic} survey strategies, and
utilise rigorous signal:noise calculations to calculate the resulting posterior
constraints on the physical properties of galaxies. We explore a broad 
range of assumed input galaxy types ($>$ 20,000 models, including extremely dusty objects) across
a wide redshift range (out to $z \simeq 12$), while
at the same time considering a realistic mix of galaxy properties based on
our current knowledge of the evolving population (as quantified through the Empirical Galaxy Generator: EGG).
While our main focus is on imaging surveys with NIRCam, spanning $\lambda_{obs} = 0.8-5.0\,\mu$m, an important goal of this work is to quantify the impact/added-value 
of: i) parallel imaging observations with MIRI at longer wavelengths, and ii) deeper supporting optical/UV imaging
with {\it HST} (potentially prior to {\it JWST} launch)
in maximising the power and robustness of a major extragalactic NIRCam survey.
We show that MIRI parallel 7.7-$\mu$m imaging is of most value for better constraining the
redshifts and stellar masses of the dustiest ($A_V > 3$) galaxies, while
deep $B$-band imaging (reaching $\simeq 28.5$\,AB mag) with ACS on
{\it HST} is vital for determining the redshifts of the large numbers of
faint/low-mass, $z < 5$ galaxies that will be detected in a deep
{\it JWST} NIRCam survey.

\end{abstract} 
\begin{keywords}
   galaxies: evolution, galaxies: formation, galaxies: high-redshift, galaxies: photometry, telescopes
\end{keywords}

% ------------------------------------------------------------------

\section{Introduction}\label{sec1:introduction}

In recent years, deep, multi-band imaging surveys have played a key role in
advancing our understanding of galaxy evolution. Deep optical surveys have been
conducted with wide-format CCD cameras on ground-based telescopes such as Subaru
(e.g. \citealp{Taniguchi:2007}, \citealp{Furusawa:2008}, \citealp{Furusawa:2016}), and with the Advanced Camera for Surveys (ACS) on the {\it Hubble Space Telescope} {\it HST} (e.g. 
\citealp{Giavalisco:2004}; \citealp{Beckwith:2006}; \citealp{Scoville:2007}). Over the past decade these studies have been extended
into the near infrared with ground-based imagers such as WFCAM on UKIRT
(\citealp{Lawrence:2007}; \citealp{Hartley:2013}), VIRCAM on VISTA
(\citealp{McCracken:2012}; \citealp{Jarvis:2013}) and HAWK-I on ESO's Very Large Telescope (VLT) (\citealp{Fontana:2014}; \citealp{Brammer:2016}), and with Wide Field Camera 3 (WFC3) on {\it HST} (e.g.
\citealp{Grogin:2011}; \citealp{Ellis:2013}; \citealp{Illingworth:2013}; \citealp{Lotz:2017}). This extension in wavelength
coverage to $\lambda \simeq 2$\,$\mu$m (or $\lambda \simeq 1.6$\,$\mu$m in the
case of {\it HST}) has been crucial in pushing back our knowledge of galaxy
luminosity functions (e.g. \citealp{McLure:2010,McLure:2013}; \citealp{Bouwens:2011,Bouwens:2015,Bouwens:2017}; \citealp{Bowler:2014, Bowler:2015}; 
\citealp{Finkelstein:2016}; \citealp{McLeod:2015, McLeod:2016}; \citealp{Ishigaki:2018}; \citealp{Oesch:2018}) 
and stellar mass functions (e.g. \citealp{Fontana:2006}; \citealp{Muzzin:2013}; \citealp{Tomczak:2014}; \citealp{Davidzon:2017}) to earlier times, 
facilitating the study of red dust-obscured objects and passive
galaxies at redshifts $z > 1$ (e.g. \citealp{Dunlop:2007,Dunlop:2017}; \citealp{Fontana:2009}; \citealp{Bourne:2017}; \citealp{Glazebrook:2017}; \citealp{Koprowski:2018}; \citealp{Merlin:2018}), 
and improving the quality of photometric redshift
information at all redshifts (e.g. \citealp{Cirasuolo:2010}; \citealp{Ilbert:2013}; \citealp{Dahlen:2013}; \citealp{Simpson:2014}; 
\citealp{Santini:2015}; \citealp{Parsa:2016}; \citealp{Laigle:2016}; \citealp{McLure:2018}). Extension to longer near/mid-infrared
wavelengths has proven more difficult, as it is essentially impossible to achieve
imaging of sufficient depth from the ground at $\lambda > 2.3$\,$\mu$m for the study of distant galaxies. As a
result the provision of useful data at $\lambda \simeq 3 - 30$\,$\mu$m has had
to rely on the much smaller {\it Spitzer Space Telescope}. The data produced by
deep surveys with the IRAC and MIPS instruments on-board {\it Spitzer} have been
undeniably impressive
(e.g. \citealp{Sanders:2007}; \citealp{Damen:2011}; \citealp{Ashby:2013,Ashby:2015,Ashby:2018}; \citealp{Labbe:2013,Labbe:2015}), and have certainly demonstrated the power of such longer-wavelength
imaging for rest-frame optical/near-infrared studies to the very earliest times (e.g. \citealp{Dunlop_Review:2013}; \citealp{Stark:2016}),
but nonetheless they are ultimately limited in angular resolution ($> 1.5$\,arcsec) and depth (to $m_{AB} \simeq 25$\,mag)
by the small (85-cm) aperture of
{\it Spitzer}.

Both the resolution and depth of near/mid-infrared imaging will be revolutionized by the {\it James Webb Space Telescope} ({\it JWST}; \citealp{Gardner:2006}), 
currently now scheduled for launch in 2021. As a result of its 6.5-m diameter primary mirror, its planned passively-cooled operation from 
Lagrangian point two (L2; 1.5$\times$10$^6$ km anti-sunward from the Earth), 
and its powerful near/mid-infrared instrumentation suite, the advances offered by {\it JWST} for the study of galaxies in the young Universe are truly spectacular. Most uniquely transformative 
are its capabilities at wavelengths $\lambda > 2$\,$\mu$m, which will never be rivalled from the ground, even in the era of 25-40m telescopes.

There are four main instruments on-board {\it JWST}:  NIRSpec (Near Infrared Spectrograph, \citealp{Posselt:2004}),
MIRI (Mid-Infrared Instrument, \citealp{Rieke_MIRI:2015}), NIRCam (Near-Infrared Camera, \citealp{Horner:2004}), 
and FGS/NIRISS (Fine Guidance System/Near-InfraRed Imager and Slitless Spectrograph, \citealp{Doyon:2012}). While the spectroscopic capabilities of {\it JWST} clearly promise enormous advances 
in our understanding of the physical properties of galaxies (especially at high redshifts), the focus of the work presented here is on planned deep imaging surveys, 
primarily with NIRCam, but also potentially involving MIRI. 

Modest programmes of NIRCam and MIRI imaging have already been approved as part of the {\it JWST} Early Release Science (ERS) programme
({\it e.g.} the Cosmic Evolution Early Realease Science (CEERS) survey; \citealp{Finkelstein:2017}) with more ambitious imaging also planned as part of the guaranteed time programs in Cycle-1 (e.g. as part of the {\it JWST} Advanced Deep Extragalactic Survey (JADES) program\footnote{\url{https://www.cosmos.esa.int/web/jwst-nirspec-gto}}\footnote{\url{https://issues.cosmos.esa.int/jwst-nirspecwiki/display/PUBLIC/Overview}}).  CEERS plans to provide 100 arcmin$^2$ of JWST imaging and spectroscopy across the majority of the Extended Growth Strip {\it HST} legacy field with approximately 37 hours of science integration time at a total cost of $\simeq$ 63 hours. CEERS utilises 10 NIRCam prime imaging pointings with 6 parallel NIRSpec spectroscopy observations and 4 MIRI imaging parallels.This results in an estimated depth capability of $m_{AB} \simeq 28-29$ mag for NIRCam and $m_{AB} \simeq 25-26$ mag in MIRI F770W. Therefore, CEERS  NIRCam imaging should enable the detection of $\simeq 50$  galaxies between $z \simeq 9 - 13$. However, the shortest wavelength filter used in CEERS is the NIRCam F115W band which places the Lyman break at around $z \simeq 9$. Therefore, by design, CEERS will heavily rely on sufficient {\it HST} data to maximise the scientific output of the resulting NIRCam dataset.

JADES  is a {\it JWST} Cycle-1 programme designed by the NIRSpec and NIRCam GTO team, aiming to observe in both the CANDELS GOODS-N and GOODS-S fields with a two-tier exposure strategy consisting of a `Deep' and `Medium' imaging survey. The `Deep' survey, centred on GOOD-S/HUDF, will reach depths of $m_{AB} \simeq 29.8$ mag across 46 arcmin$^2$ with no MIRI imaging parallels. The shallower `Medium' survey will cover a wider area of $\simeq 190$\,arcmin$^2$ to a limiting magnitude of $m_{AB} \simeq 28.8$. Here, there will be a total of $\simeq$ 14 arcmin$^2$ of parallel MIRI imaging in the F770W filter, with approximately 8 arcmin$^2$ reaching a limiting magnitude of $m_{AB} \simeq 26.7$ mag. Like CEEERS, the dataset produced will require sufficient {\it HST} optical data to robustly select Lyman break galaxies at $z < 7$. However, as outlined in \cite{Finkelstein:2015}, the community has ambitions to go beyond these programs and to engage in more ambitious NIRCam+MIRI surveys increasing the contiguous area coverage and depth across all the CANDELS fields (COSMOS and UDS) in the early years of {\it JWST} 
operations. This will enable the community to discover and study hitherto undetected populations of galaxies (at extreme redshifts, low masses and/or extreme dust obscurations), to better constrain the redshifts, stellar masses, and rest-frame optical morphologies of known high-redshift galaxies, and also to select future targets for {\it JWST}.
spectroscopic follow-up within the lifetime of the mission. One attractive feature of larger area ($> 100$\,arcmin$^2$) NIRCam mapping is that parallel-mode {\it JWST} observing
can be used to drag MIRI (which lies $\simeq 6$\,arcmin distant from NIRCam in the telescope focal plane) over regions already mapped with NIRCam, to produce homogeneous imaging of consistent depth
across the full wavelength range $0.8 < \lambda < 20$\,$\mu$m, exploiting to the full {\it JWST's} near-to-mid infrared imaging capabilities. 

Such {\it JWST} imaging surveys promise huge advances in our understanding of cosmic history, but, given the substantial investment in observing time required for large area mapping with {\it JWST}, 
and the range of possible observing strategies, their detailed design demands to be based on careful analysis. Notwithstanding the obvious fact that deep {\it JWST} imaging 
will open up new discovery space, enough is now known about the expected performance of the telescope and instrumentation, and the properties of the known galaxy population, to merit
a proper quantitative evaluation of the trade-offs between different choices of NIRCam filter combinations, and survey depth versus area. There are two key related aspects to consider, 
which are the primary focus of this paper, namely i) the added value of MIRI parallel observations (which while nominally `for free' can set significant constraints on survey layout, scheduling etc)
and, ii) the depth/quality of {\it optical} data in the chosen survey fields required to maximise the power of the investment planned with {\it JWST}.

The approach taken here, considering fully realistic photometry, and deriving full posterior probability distributions for derived galaxy physical parameters under different observational strategies differs somewhat from that of other published studies aimed at providing extragalactic predictions for {\it JWST}.
In recent years, a number of such `{\it JWST} prediction' papers have appeared, typically analysing mock {\it JWST} catalogues built using semi-analytic models, or numerical cosmological hydrodynamical simulations of galaxy formation and evolution 
(e.g. \textsc{FiBY} \citealp{Paardekooper:2013}, BlueTides \citealp{Feng:2016}; \citealp{Wilkins:2017}, GALFORM \citealp{Lacey:2016}, \citealp{Barrow:2017}, \citealp{Lovell:2018}, \citealp{Yung:2019}). 

\begin{figure*}
\begin{center}
\includegraphics[angle=0,scale=0.49]{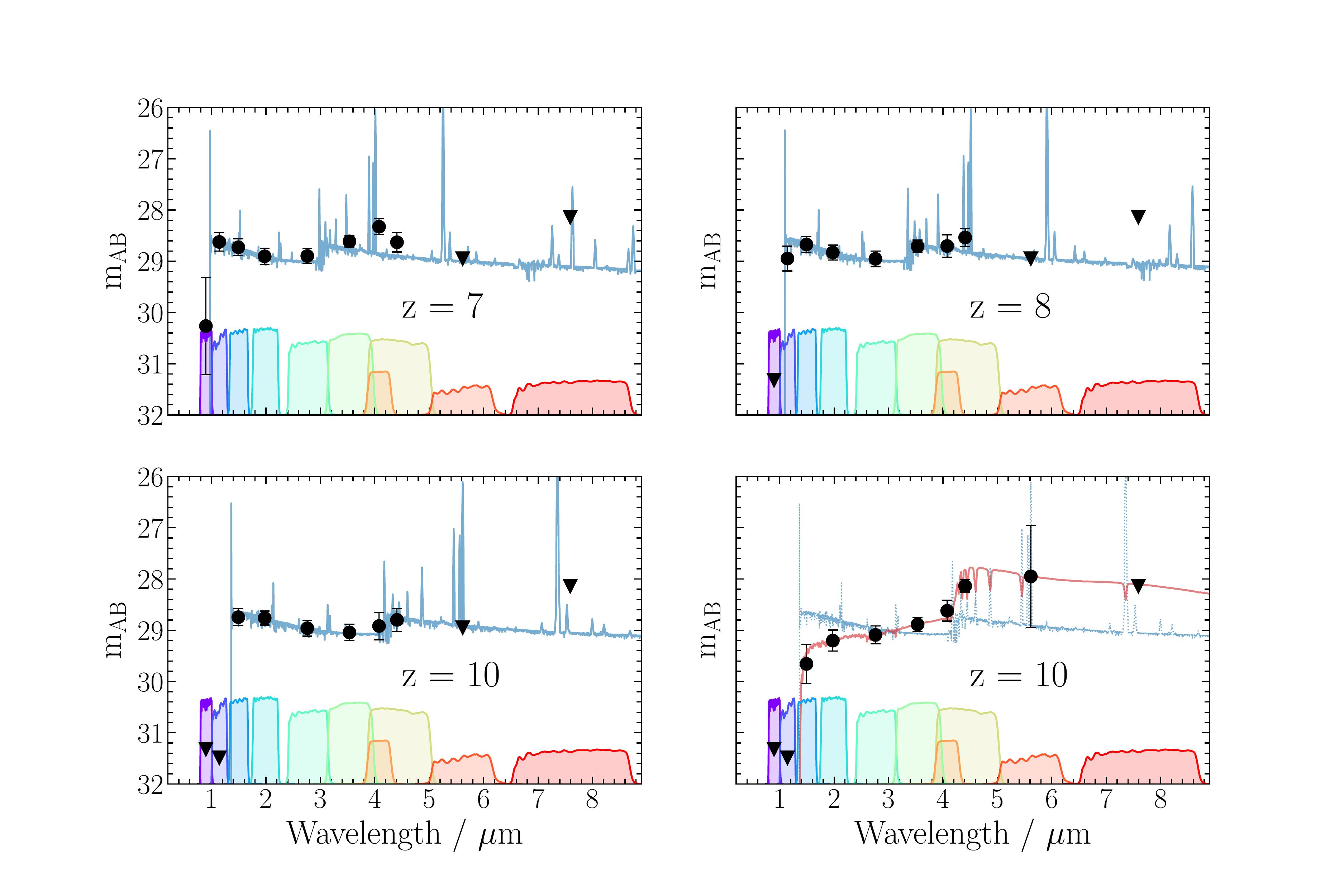}
\end{center}
\caption{The power of multi-band NIRCam + MIRI
  imaging for the discovery and study of high-redshift 
  galaxies. 
  The blue SED shown at $z = 7,8,10$ is that of a FiBY
  (\citealp{Paardekooper:2013}; \citealp{Cullen:2017}) simulated star-forming
  galaxy with $Z/{\rm Z_{\odot}} \simeq 1/3$,
  normalised to $m_{\rm 3\mu m} = 29$\,mag, with the data-points showing the simulated photometry (and associated uncertainty) assuming 1-hr 
exposures per NIRCam filter. 
  As well as clear delineation of the Lyman-break, the impact of the
  strong rest-frame optical emission lines (produced by sub-solar metallicity galaxies) on the photometry is clear;
  F444W$-$F410M colour is invaluable for redshift refinement, and provides estimates of emission-line equivalent widths, and resulting corrections to stellar mass. The 4th 
  panel shows an alternative (red) SED at $z \simeq 10$, produced by a starburst at $z \simeq 15$ ($\simeq 200$\,Myr old), demonstrating that NIRCam + MIRI imaging 
  can also provide key information on earlier star-formation activity.}\label{fig:highz} 
\end{figure*}

As a recent example, \cite{Cowley:2018} built a mock galaxy catalogue based on the GALFORM semi-analytic model, and used this to predict galaxy number counts, redshift distributions and luminosity functions in all broadband filters available for NIRCam and MIRI. They predict that assuming a 10$^4$s exposure, very few galaxies will be detectable per field-of-view beyond $z > 10$ and $z > 6$ with NIRCam and MIRI, respectively. They also estimate the size of the galaxies in each photometric band in order to understand the resolving capabilities of NIRCam and MIRI at high redshift. \cite{Williams:2018} made similar predictions for NIRCam only, assuming the observational prescriptions of the {\it JWST} Advanced Deep Extragalactic Survey (JADES) GTO program. 
They show that they expect to detect thousands of galaxies at $z > 6$ with tens at $z > 10$ across the entirety of the JADES NIRCam imaging (236 arcmin$^2$) as well as constraining the evolution of the galaxy UV luminosity function at $z > 8$. However, the goal of these studies was not to investigate the accuracy and reliability with which the {\it JWST} data would enable simple galaxy properties, such as redshift and stellar mass to be robustly reclaimed. This means that the predicted results are inevitably somewhat unrealistic, and also precludes quantitative comparison of the power of alternative observing strategies, and the importance of supporting data outside the wavelength range of NIRCam.

One study which tested how reliably galaxy properties can be reclaimed under different filter combinations of {\it JWST} photometry is that of \citet{Bisigello:2016, Bisigello:2017}.
First focussing on photometric redshifts, \citet{Bisigello:2016} created a mock galaxy catalogue, added
appropriate observational noise, and then used spectral energy distribution (SED) fitting in order to estimate the physical properties of each `observed' galaxy. They concluded 
that, in isolation, NIRCam data struggle to deliver acceptable photometric redshifts in certain redshift ranges, and they attempted to explore which filter combinations offered the best 
prospect of overcoming (or at least ameliorating) this problem. Their main conclusion is that, while deep optical data at $\lambda < 0.6$\,$\mu$m from {\it HST} is desirable, the lack of such 
short-wavelength information can be mitigated somewhat by the use of MIRI photometric observations in the F560W and F770W filters. In particular, they reported that the 
addition of such MIRI imaging to a NIRCam survey can reduce the photometric redshift Catastrophic Outlier Rate (COR, defined as $|\Delta z|$ $>$ 0.15, where $|\Delta z|$ = $|(z_{phot} - z_{input})/(1+z_{input})|$, by a significant fraction both with and without the addition of {\it HST} data. 
However, in coming to this conclusion, the authors assumed that a MIRI detection at 28 AB mag would have a comparable S/N as a NIRCam detection 
at 29 AB mag in the F150W filter. This is an optimistic (and, in practice, unrealistic) assumption for an imaging survey, since it requires MIRI integrations $\simeq$\,50 times longer than the NIRCam integrations in this scenario. Therefore, this is not appropriate to the anticipated real-world case described above for MIRI imaging being obtained in parallel to NIRCam imaging, and would require a NIRCam+MIRI survey design that committed $>$ 95\% of the available integration time to MIRI imaging. In a second paper, \citet{Bisigello:2017} then focussed on exploring the optimum combination of NIRCam and MIRI
filters for reclaiming the physical properties of a sample of 1542 simulated galaxies in the high redshift range, $z = 7 - 10$. Again they reported that the addition of MIRI data was very valuable, especially
for improving stellar mass and specific star-formation rate (sSFR)
estimation at $z \simeq 10$. However, again these calculations are
not really realistic for a {\it JWST} imaging survey,
as a signal:noise ratio of
$S/N = 10$ is assumed for a galaxy with $m_{AB} \simeq 28$\,mag in the F560W filter which, even with a low background level, requires $\simeq 200$\,hr of integration per MIRI pointing. 

The aim of the study presented here was therefore to: i) consider a set of fully schedulable and realistic observing scenarios, ii) explore a wide range of assumed input galaxy types ($> 20,000$ alternative model galaxies),
iii) perform detailed signal:noise calculations to perturb the resulting predicted photometry, and iv) use state-of-the-art Bayesian methods 
to assess the precision and accuracy with which basic galaxy physical parameters can be reclaimed. Armed with such machinery we are able to accurately assess the power of alternative observing strategies (both for the galaxy population in general, and for special galaxy subclasses of potential interest) and, specifically, the importance of complementing NIRCam imaging with MIRI observations and/or
additional {\it HST} optical imaging than already exists in the likely target {\it JWST} survey fields.

The structure of this paper is as follows. In Section \ref{sec2:instrumentation} we summarize the key properties of the NIRCam and MIRI imagers,
and explain the rationale behind our adopted NIRCam filter combination and observing strategy for the discovery and study of high-redshift galaxies.
Next, in Section \ref{sec3:sim_galaxies}, we describe the methods used to produce simulated galaxies. Here we aimed both to sample a wide range of possible galaxy properties (using the models of
Bruzual \& Charlot (2003), with star-formation histories informed by the First Billion Years (\textsc{FiBY}) simulations; \citealp{Paardekooper:2013}; \citealp{Cullen:2017}), and to consider
realistic models of the evolving galaxy population (based on current data, using The Empirical Galaxy Generator (EGG; \citealp{Corentin:2017}). Then, in Section \ref{sec4:sim_obs}, we describe the process by which we synthesized the predicted 
{\it JWST} photometry for the simulated galaxies, considering realistic NIRCam+MIRI observing strategies. In Section \ref{sec5:reclaim}, we explain how we utilised
the Bayesian SED-fitting code \textsc{Bagpipes} \citep{Carnall:2018} to reclaim the physical parameters of the simulated galaxies from the {\it JWST} photometry, and explore the accuracy with which this can be achieved both with and without the addition of MIRI and deeper {\it HST} photometry to the core NIRCam dataset. Finally, in Section \ref{sec6:sum_conclusion}, we give a summary of our main
conclusions. Throughout the paper magnitudes are quoted in the AB system (\citealp{Oke:1974}; \citealp{Oke:1983}) and we assume a flat cosmology with $\Omega_M = 0.3$, $\Omega_{\Lambda} = 0.7$, and $H_0 = 70\,{\rm km\,s^{-1}\,Mpc^{-1}}$.

\section{Imaging high-redshift galaxies with JWST}\label{sec2:instrumentation}

\subsection{The JWST imagers}

Full details of the {\it JWST} instrumentation are publicly available online\footnote{\url{https://jwst.stsci.edu/instrumentation}},\footnote{\url{https://www.nasa.gov/mission_pages/webb/instruments/index.html}},\footnote{\url{https://jwst-docs.stsci.edu/display/JTI}}.  The key features of NIRCam and MIRI can be summarized as follows.

The Near-Infrared Camera (NIRCam, \citealp{Horner:2004}) is the primary imager onboard {\it JWST} with a wavelength range of $0.6< \lambda < 5 \mu$m. There are two identical modules, A and B, where the field of view (FOV) is split into a short-wavelength channel and a long-wavelength channel with co-aligned FOV's (2.2 $\times$ 2.2\,arcmin). The short-wavelength channel 
($\lambda~=~0.6~-~2.3$\,$\mu$m) has four 2040$\times$2040 detector arrays with a pixel scale of 0.032\,arcsec per pixel in each module. The long-wavelength channel ($\lambda~=~2.4~-~5$\,$\mu$m) has only one 2040$\times$2040 detector array in each module with a pixel scale of 0.065\,arcsec per pixel. The short-wavelength channel has twelve filters in total; five wide-band (R $\sim$ 4) filters, four medium-band (R $\sim$ 10) filters, and three narrow-band (R~$\sim$~100) filters. The long-wavelength channel consists of three wide-band, eight medium-band, and four narrow-band filters. In total, there are therefore 8 available broad-band filters: F070W, F090W, F115W, F150W, F200W, F277W, F356W and F444W. With such filters, the key spectral feature 
of extreme redshift galaxies, the Lyman break at $\lambda_{rest} = 1216$\AA, can in principle be detected over the redshift range $4.2 < z < 40$, while the 4000\AA/Balmer break 
feature is identifiable over the redshift range $0.6 < z < 11.3$.

The Mid Infrared Imager (MIRI, \citealp{Bouchet:2015}) is a combination of multiple instruments providing observing capabilities over a vast range of MIR wavelengths ($\lambda~=~5~-~27$\,$\mu$m). In broad-band imaging mode, nine photometric broad bands are available, with resolution capabilities of R $\sim$ 5 (F560W, F770W, F1000W, F1130W, F1280W, F1500W, F1800W, F2100W, F2550W). Along with broad-band imaging, this module contains instruments that perform coronography and low-resolution spectroscopy (LRS) \citep{Kendrew:2015}. The pixel scale of the imager is 0.1\,arcsec per pixel with a total FOV of 74$\times$113\,arcsec devoted to imaging alone.  MIRI's sensitivity, combined with its long wavelength baseline, will provide unrivalled mid-infrared data. The wavelength range accessible with MIRI becomes very important for high-redshift galaxies (especially at $z >6$), where the rest-frame optical continuum and several of the key nebular emission lines become redshifted longward of $\lambda \simeq 5$\,$\mu$m. MIRI's first two bands are by far the most sensitive, with the F560W band sitting very close to the end of the NIRCam wavelength range. 

\subsection{JWST imaging strategy for high-redshift galaxy surveys}

Although WFC3/IR enabled {\it HST} to probe beyond $z \simeq 7$ into the first $\simeq$ Gyr of cosmic time,
the isolation of secure samples of $z > 7$ galaxies has been severely hampered by the curtailment of
{\it HST} wavelength coverage at $\lambda_{\rm obs} < 1.7$\,$\mu$m. Robust redshift information benefits greatly not only from identification
of the Lyman-break at $\lambda_{\rm rest} = 1216$\,\AA, but also from extended/high-quality wavelength coverage redward of the 
break to exclude lower-redshift red/dusty interlopers. Here, {\it JWST} will be transformative; as we show in Fig.\,\ref{fig:highz},
an 8-filter NIRCam approach can not only securely identify Lyman-break galaxies out to the highest redshifts, but can
also provide powerful SED information (e.g. revealing the presence
of rest-frame optical emission-lines or a Balmer break).

As can be seen from Fig.\,\ref{fig:highz}, 7 NIRCam broad-band filters (F090W, F115W, F150W, F200W, F277W, F356W, F444W) 
provide complete (and independent) coverage of the observed spectral range $\lambda = 0.8 - 5$\,$\mu$m. 
We have decided to adopt this as our baseline NIRCam filter set for the simulations presented here, with inclusion of the F090W filter (excluded in the CEERS ERS program) vital 
for delivering samples of galaxies at $z = 7-8$. We do not include imaging in the F070W filter, because i) this is the least sensitive short-wavelength channel broad-band filter for NIRCam imaging, ii) this wavelength
range is least unique to {\it JWST}, and iii) the image quality in this filter is not expected to be significantly better than achieved with {\it HST} at comparable wavelengths.

We also assume equal exposure times in all 7 of these broad bands. There are two reasons for this. First, since imaging is 
performed simultaneously in a short-wavelength
channel filter and a long-wavelength channel filter, it is not straightforward/practical to fine-tune the relative exposure times (e.g. increasing systematically with wavelength). Second, the imaging sensitivity expected in each of these filters (in terms of AB mag or $\mu$Jy) is comparable (to within 0.5 mag, with the exception of F444W\footnote{\url{https://jwst-docs.stsci.edu/display/JTI/NIRCam+Imaging}}), and thus already well
matched to the consistent detection of high-redshift star-forming galaxies, which are known to have a fairly `flat' (i.e. $\beta = -2$, where $f_{\lambda} \propto \lambda^{\beta}$) rest-frame 
UV continuum above the Lyman-break, as shown in Fig.\,\ref{fig:highz}.

Having decided on a combination of 4 broad-band filters in the short-wavelength NIRCam channel, and all 3 available broad-band filters in the long-wavelength
channel, it only remains to decide on the choice of a fourth long-wavelength channel filter. The obvious options are additional integration in one of the broad-band filters
already included (i.e. F277W, F356W, F444W) or inclusion of imaging in a different long-wavelength channel filter. 
In practice, this becomes a choice between more integration in F444W (given it is the 
least sensitive of the broad bands in the long-wavelength channel) or inclusion of a long-wavelength channel medium-band filter in a comparable wavelength regime. 
For constraining continuum SEDs, experiments indicate that this choice is unimportant as the relative sensitivity capabilities and effective wavelengths are somewhat similar for the F444W and F410M NIRCam filters \footnote{ Although the F444W band is slightly more sensitive than F410M band for most observations, the background is steeply rising between 4 and 5 $\mu m$ (this is the Wien tail of the thermal zodiacal dust spectrum). The F444W filter goes out to almost 5 $\mu$m, so it takes the brunt of this rise. This significantly counteracts the sensitivity gains that would otherwise be obtained from the wider bandpass. Because F444W is more sensitive to the background than F410M, any observation with long exposure time becomes background-dominated in F444W, and thus the flat-field residual noise on the background becomes significant for background-dominated calculations.}. However, here we choose to utilise the medium-band F410M filter (as again shown in Fig.\,\ref{fig:highz}) 
because of its potential to help distinguish between a spectral break at the red end of the NIRCam imaging, or the presence of strong emission lines 
within the broad band (similar baseline filter set in the JADES program). In this regard, the combination of F444W and the medium-band F410M filter can be particularly powerful.
 It is now well-established that [OIII]
lines of extreme equivalent widths (EW$_0$) become more prevalent with increasing redshift and/or decreasing metallicity (\citealp{Nakajima:2016}),
and, for brighter objects, have already proved detectable in broad-band {\it Spitzer} IRAC photometry (\citealp{Smit:2015}). Not
only can such emission-line signatures provide invaluable additional redshift information, but they also offer one
of the few ways of estimating the escape fraction of ionizing photons from young galaxies
in the reionization epoch (\citealp{Stark:2016}).

Given this NIRCam imaging strategy, one of the key issues we aim to explore in this work is the added value of MIRI imaging for high-redshift galaxy studies.
In principle this would appear to be extremely valuable. The wavelength coverage of NIRCam and MIRI compliment each other nicely, and while surveys with NIRCam alone will 
effectively encompass the 4000\AA/Balmer break displayed by galaxies across the broad redshift range $1 < z < 10$, at $z > 4$, the wavelength range sampled by 
MIRI better samples the light from the evolved stellar populations which dominate the stellar masses of galaxies. Moreover, the enhanced wavelength baseline offered
by the addition of MIRI imaging can be of value for distinguishing between quiescent and dusty star-forming galaxies, and hence establishing 
reliable galaxy masses for red sources and completing our census of the stellar-mass density at early times. 

The appeal of adding MIRI imaging is, at first sight, further enhanced by the fact that it can be obtained in parallel with NIRCam imaging. MIRI lies $\simeq 6$\,arcmin from NIRCam 
in the {\it JWST} focal plane, and thus images a different part of the sky. However, for moderately wide-area surveys (e.g. of GOODS/CANDELS fields $> 100$\,arcmin$^2$) the observations
can be designed to drag the MIRI parallel imaging over the majority of the area imaged with NIRCam, and thus the MIRI imaging can be obtained `for free'. However, such a strategy 
means that the MIRI integration times are, necessarily, limited to be comparable to the NIRCam integration times. One could of course choose to keep observing in a single 
MIRI filter while the observations at a given location cycle through all of the NIRCam filters described above, but that still sets a maximum MIRI integration time of $\simeq 4 \times$ the NIRCam 
integration time per filter. This is then potentially problematic when trying to close the gap between the detection limits of NIRCam and MIRI. Even the two most sensitive MIRI filters at wavelengths immediately longward of those probed by NIRCam, F560W and F770W, 
are $\simeq 10 \times$ less sensitive than NIRCam F444W imaging. The sensitivity rapidly degrades further with increasing wavelength in the 
MIRI channels, largely due to the increased background noise from the JWST insturments themselves\footnote{\url{https://jwst-docs.stsci.edu/display/JTI/MIRI+Sensitivity}}. This means that realistic calculations need to be performed to assess the true added value of MIRI photometry to deep multi-band NIRCam imaging, including 
a full simulation of the accuracy with which the physical parameters of galaxies over a range of masses, redshifts and spectral types can be reclaimed with and without the addition 
of MIRI parallel imaging.

In the simulations presented here we assume that all the available parallel imaging integration time is devoted to 7.7-$\mu$m MIRI imaging (i.e. F770W) because
i) the sensitivites offered by F560W and F770W imaging are comparable, and ii) 7.7-$\mu$m imaging most significantly extends the wavelength baseline covered by NIRCam. We stress that 
important extragalactic science can undoubtedly be undertaken using the longer-wavelength MIRI filters, but the aim here is to test the extent to which MIRI imaging in the most sensitive MIRI channels, taken in parallel with deep NIRCam imaging, can really contribute to our knowledge of the tens of thousands of previously undiscovered galaxies which will be revealed by 
deep NIRCam imaging surveys.

Finally, while Fig.\,\ref{fig:highz} shows that at redshifts $z > 7$ essentially all of the transmitted galaxy SED lies longward of $\lambda \simeq 1$\,$\mu$m, and is thus fully sampled by {\it JWST}, 
it remains the case that the vast majority of the new galaxies which will be discovered in deep {\it JWST} surveys will lie at more modest redshifts. This means that shorter-wavelength 
(optical) observations can still be of value in constraining the galaxy SEDs, and raises the question of whether the exisiting optical (e.g. {\it HST} ACS) imaging in likely target
{\it JWST} survey fields
is adequate to enable full exploitation of deep {\it JWST} imaging data. This is the second key strategic issue that we explore here through our galaxy simulation work.

\begin{figure}
  \begin{center}
\includegraphics[scale=0.29]{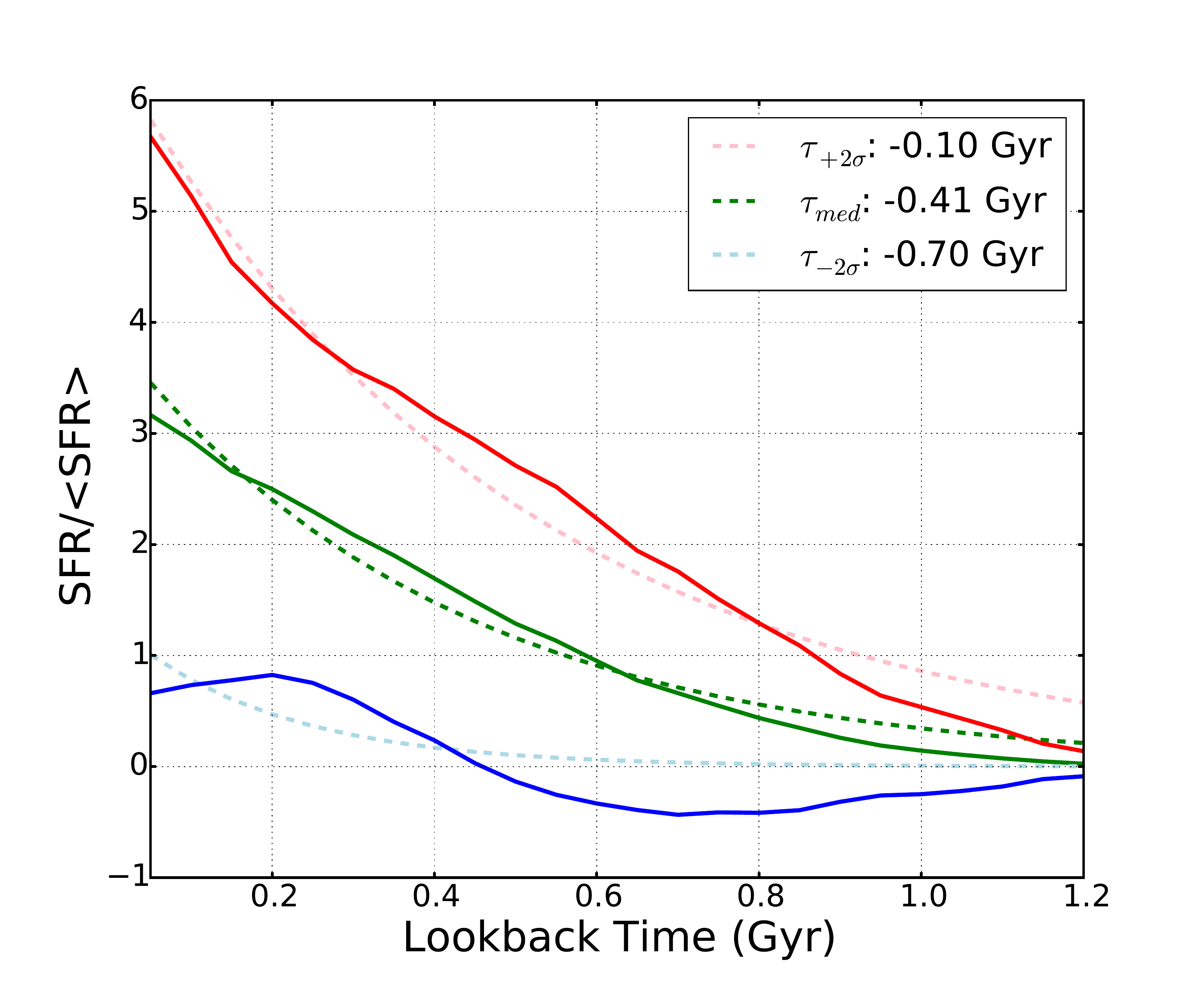}
  \end{center}
\caption{Polynomial fits to the median \textsc{FiBY} SFHs for $z = 4$ galaxies. The solid coloured lines represent the median (green) and $\pm$2$\sigma$ (blue and red) normalised SFHs across the mass range $10^{8} - 10^{10}\,{\rm M_{\odot}}$. The dashed lines show the minimum-$\chi^2$ polynomial fits to the solid lines. The resulting extracted rising star-formation $\tau$ values 
are displayed in the top-right corner. These are the $\tau$ values adopted to build our suite of exponentially-increasing BC03 galaxies}\label{figure:tau_fit}
\end{figure}

\begin{table}
\caption {Physical parameter values used to build the suite of model galaxies based on the BC03 templates for our simulated galaxies at $z = 3 - 9$. These values were used for all of the SFH prescriptions detailed in Table\, \ref{table:SFH}.}\label{table:BC03_Model}
\begin{center}
    \begin{tabular} { l | l | l }
      \hline
      \textbf{Property} & \textbf{Parameter}  & \textbf{Value(s)}  \\
      \hline
            Stellar mass & $log_{10}(M_*/{\rm M_{\odot}})$  & 8, 9, 10 \\
          Metallicity & $Z$ / ${\rm Z_{\odot}}$ & 0.2, 1 \\
          Redshift & z & 3, 3.5, 4 ... 8, 8.5, 9\\ 
          Dust obscuration & $A_{V}$ / mag & 0, 0.2, 0.4, 0.8 ... 6.4 \\
          Ionisation parameter & $\log_{10} U$ & $-$3 \\
          Age of the Birth Cloud & $T_{bc}$ / Gyr & 0.01 \\
    
    \end{tabular}
  \end{center}

\end{table}

\section{Simulated Galaxies}\label{sec3:sim_galaxies}

In this section, we describe the two suites of galaxy models we constructed to explore the potential power of {\it JWST} imaging surveys. First we constructed an extensive (deliberately not necessarily representative) range 
of galaxy SEDs utilising the models of \cite{B&C:2003} (BC03). As described in detail below (subsection \ref{section3:bc03}), we explored the impact of a wide range of physical properties, in part to enable 
a search for those types of (possibly extreme) galaxies for which MIRI data could have the most significant effect on reclaimed galaxy parameter estimates. Second, as described in subsection \ref{section3:EGG}, we constructed a more 
representative (at least given current knowledge) distribution of galaxy SEDs using the Empirical Galaxy Generator (EGG) simulations (\citealp{Corentin:2017}), in an attempt to quantify the actual number/redshift-distribution of galaxies which will be uncovered by deep NIRCam surveys, and assess the relative prevalence of the likely subsets of galaxies for which additional MIRI or {\it HST} photometry is predicted to be of most value.   

\subsection{BC03 Models} \label{section3:bc03}

Using the population synthesis models of \citet{B&C:2003}, we created a set of 21,216 galaxy SEDs covering a vast range of specified physical parameters. We built this suite of galaxies using the model generation component of the spectral fitting code Bayesian Analysis of Galaxies for Physical Inference and Parameter Estimation (\textsc{Bagpipes}; \citealp{Carnall:2018}). We spread the BC03 templates across 13 evenly-spaced redshift bins ($z = 3$, 3.5, 4 ... 8, 8.5, 9), and, at each redshift, constructed galaxies at three stellar masses of potential interest ($M_{*} = 10^8, 10^9, 10^{10}\,{\rm M_{\odot}}$). At each redshift and stellar mass, we created model galaxies with both solar or 1/5 solar metallicity, explored 34 alternative star-formation histories (SFH) 
(16 exponentially-declining models, 12 exponentially-increasing models, 3 burst+constant star-formation models, and 3 constant+burst models), and considered a range of dust obscuration
extending up to $A_V \simeq 6$, applying dust attenuation using the Calzetti reddening law (\citealp{Calzetti:1994}; \citealp{Calzetti:2000}). The effect of the intergalactic medium (IGM) attenuation is also included in these models, following the prescription of \citet{Inoue:2014}. The BC03 templates do not include nebular emission, however, \textsc{Bagpipes} has fully integrated the latest version of the
{\sc cloudy} photoionization code \citep{Ferland:2017}. The logarithm of the ionisation parameter ($U$, defined as the ratio of ionising photons to neutral particles) combined with the input metallicity determines the level of nebular emission applied to the BC03 templates. In this work, we fix the logarithm of the ionization parameter to $\log_{10}(U) = -3$. More information on the implementation of these processes is detailed in \cite{Carnall:2018}.

\begin{table}
  \begin{center}
    \begin{tabular} { l | l | l }
      \hline
      \textbf{Form of SFH} & \textbf{Parameter}  & \textbf{Value(s)}  \\
      \hline
       \textbf{Exponentially Decreasing} && \\
          Timescale &$\tau$ / Gyr & 0.05, 0.25, 1.25, 6.25  \\
          Age of SFH &$T$ / Gyr & 0.02, 0.1, 0.4, $z$-lim Age \\
          & &\\
           \textbf{Exponentially Increasing} && \\
          Timescale & $\tau$ / Gyr & $-$0.1, $-$0.4, $-$0.7  \\
          Age of SFH &$T$ / Gyr & 0.02, 0.1, 0.4, $z$-lim Age \\
          & & \\
       \textbf{Burst then Constant} && \\
         Fractional mass in SF burst &$M_{b}$ / $M_{c}$ & 1/3, 1/2, 2/3  \\
          & & \\
        \textbf{Constant then Burst} & &\\
          Fractional mass in const. SF &$M_{c} / M_{b}$ & 1/3, 1/2, 2/3  \\
 
      \end{tabular}
  \end{center}
  \caption{SFH parameter values used to create the suite of SEDs based on the BC03 templates and applied to galaxies with the chosen physical parameters summarized in Table\, \ref{table:BC03_Model}.}\label{table:SFH}
\end{table}

At all redshifts the maximum galaxy age is limited by the age of the Universe at that redshift.
The burst+constant SFH models represent galaxies with an initial burst of star-formation activity at the epoch of formation, followed by a constant build-up of additional stellar mass over the remaining period of time governed by the defined age of the galaxy. In contrast, the constant+burst models are defined by an intitial extended period of constant star formation, followed by a short, sharp burst immediately before the epoch corresponding to the redshift of the observed galaxy. Each of these models is split into 3 separate mass prescriptions, with either 1/3, 1/2 or 2/3 of the total stellar mass built up in either the burst or constant component of the SFH. The exponentially-declining ($\tau$) models are, as is standard practice, defined by a range of e-folding timescales.
and ages (four of each). However, while such models may be appropriate for relatively mature galaxies in the late-time Universe, a range of observations and simulations suggest that 
rising star-formation histories are more plausible for growing galaxies in the young high-redshift Universe. In an attempt to ensure physically-realistic rising star-formation histories, we estimated appropriate $\tau$ values for exponentially-rising models via fits to the SFHs of typical young galaxies in the  First Billion Years (\textsc{FiBY}) simulations (\citealp{Paardekooper:2013}; Khochfar et al. in preparation), as shown in Fig.\,\ref{figure:tau_fit}. The \textsc{FiBY} simulation suite is a set of high-resolution cosmological hydrodynamics simulations using a modified version of the GADGET code utilised in the Overwhelmingly Large Simulations (OWLS) project \citep{Schaye:2010}. These simulations reproduce the stellar-mass function and star-formation rate of galaxies at $z > 6$ and at the same time also recover the trends in the metallicity evolution of galaxies. We extracted the median SFH of all galaxies within the stellar mass range, $10^{8} - 10^{10}\,{\rm M_{\odot}}$, and performed 
a simple polynomial (minimum-$\chi^2$) fit to the median and $\pm 2\sigma$ SFH for each redshift bin in the \textsc{FiBY} simulations. Fig.\,\ref{figure:tau_fit}, shows these fits along with the calculated $\tau$ values for the $z =4$ galaxies. The solid coloured lines represent the median and $\pm 2\sigma$ normalised SFHs extracted from the \textsc{FiBY} simulations, with the dashed lines showing the polynomial fits. We found that the median (green line) and $\pm 2\sigma$ (red and blue, respectively) $\tau$ values ($\tau = -0.1, -0.4, -0.7$\,Gyr) encompass all calculated $\tau$ values up to redshifts $z > 9$ in the \textsc{FiBY} simulations. They are also consistent with the values reported by \cite{Finlator:2011}, who found that $\tau$ should vary over the range $50-350$\,Myr within the redshift range $6 < z < 12$.   

\begin{table*}
\caption{A summary of the key features of the galaxy samples predicted (by EGG) to be detected with {\it JWST} in the the COSMOS-CANDELS and UDS-CANDELS
   fields, assuming a {\it JWST} observing time committment of $\simeq 100$\,hr to each field (see Section \,\ref{section4:Obs})}\label{table:z_stat}
 \begin{tabular}{c | c | c | c | c | c | c | c }
  \hline\hline
  Field & Area & Mass Range & Median Mass & Redshift range & Median $z$ & SFG-QG Ratio & No. of Galaxies\\
    & /arcmin$^2$ & log$_{10} (M_{\star}/{\rm M_{\odot}})$ & log$_{10} (M_{\star}/{\rm M_{\odot}})$ & & & \\
    \hline
   COSMOS & $\simeq$150 & $5.02 - 11.90$ & 8.45 & 0.050 $< z < $ 11.69  & 2.019 & 11.10  & 51,879 \\[2pt]
   UDS & $\simeq$250 & $5.11 - 11.86$ & 8.48 & 0.050 $< z < $ 10.92  & 1.969 & 10.97 & 69,766 \\[2pt]
  \hline
 \end{tabular}
\end{table*}

The complete list of adopted values is summarized in Tables\,\ref{table:BC03_Model} and\, \ref{table:SFH}, where Table\, \ref{table:SFH} summarizes
the SFH parameters. Our simulated galaxy sample contains 9,984 exponentially-decreasing, 7,488 exponentially-increasing, 1,872 burst+constant and 1,872 constant+burst model SEDs distributed 
equally across the redshift range $3 < z <  9$ and the stellar-mass range $10^{8}, 10^{9}, 10^{10}\,{\rm M_{\odot}}$.

\subsection{EGG}\label{section3:EGG}

\begin{figure*}

\centering
   \includegraphics[scale=0.3]{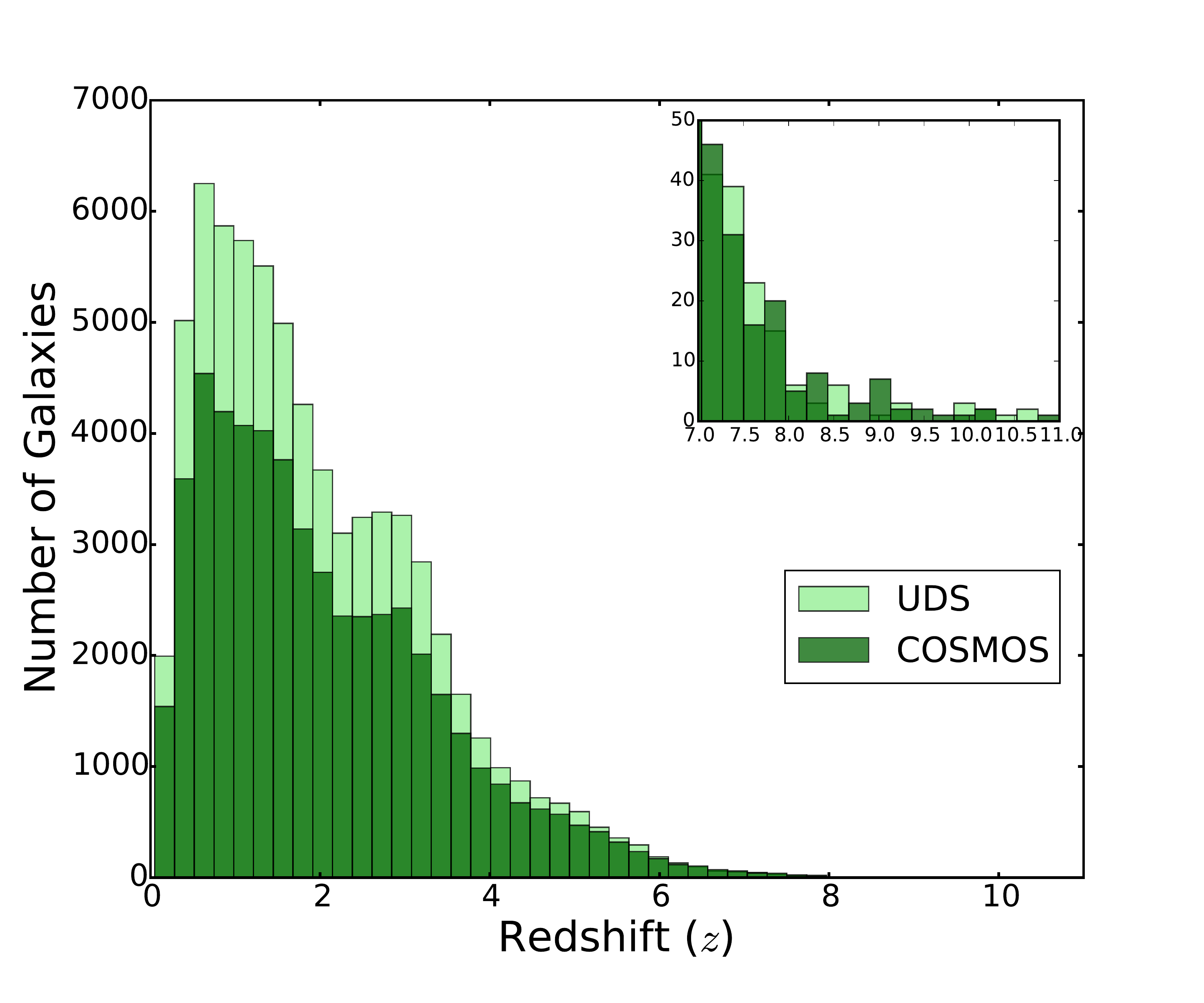}\includegraphics[scale=0.3]{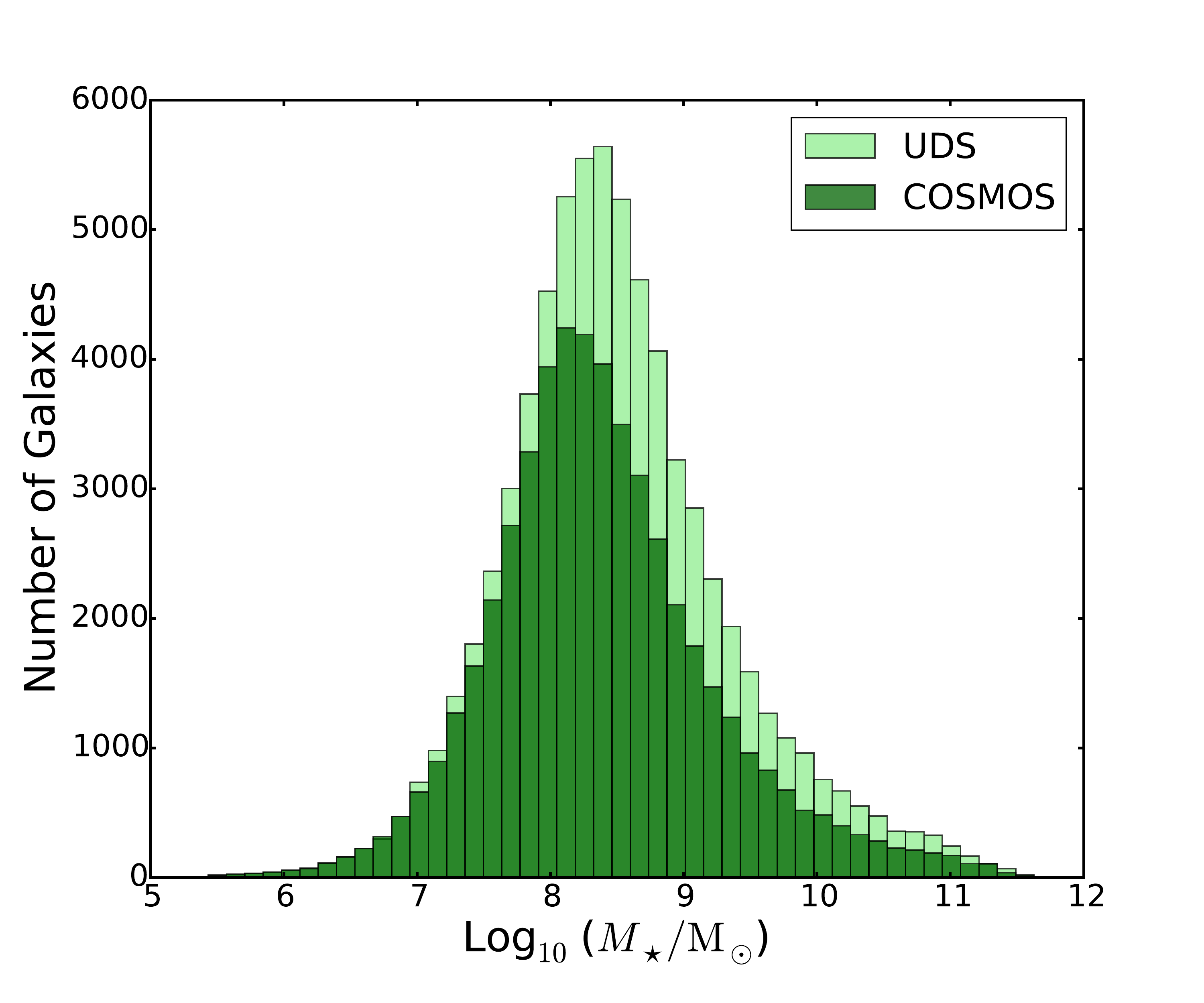}
   \caption{The redshift and stellar-mass distributions of our finalised EGG mock galaxy catalogues of {\it JWST}-detected galaxies
     in the COSMOS-CANDELS and UDS-CANDELS survey fields, observed under a realistic observing strategy for {\it JWST} (see Section\,\ref{section4:Obs}).
     The inset in the first panel provides a zoom-in of the anticipated detectable galaxy samples at $z > 7$.}
         \label{UDS_COS_Dist}
\end{figure*}

\begin{table*}
 \caption{Calculated typical {\it JWST} 5$\sigma$ point-source depths in each filter for our strawman NIRCam+MIRI survey observing strategy.}\label{tbl:filter_depths}
 \begin{tabular}{l | c | c | c | c | c | c | c | c | c }

 \hline\hline
  \textbf{Survey Field)} & F090W & F115W & F150W & F200W & F277W & F356W & F410M & F444W & F770W  \\
  \hline
  \textbf{COSMOS-CANDELS} & 28.96 & 29.23 & 29.43 & 29.51 & 29.47 & 29.42 & 28.65 & 28.93 & 25.91 \\
  \textbf{UDS-CANDELS} & 28.58 & 28.84  & 29.05 & 29.09 & 29.08 & 29.01 & 28.20 & 28.44 & 25.96 \\
  \hline
 \end{tabular}
\end{table*}

The Empirical Galaxy Generator (EGG v.1.1.0; \citealp{Corentin:2017}), is a tool that can generate large mock galaxy catalogues with realistic flux
densities. These catalogues are built using empirical prescriptions rather than deriving galaxy properties from physical first principles or
simulations, and seek to provide the most accurate match to current observational
data for galaxies in the redshift range $0 < z < 7$. EGG takes the approach of assuming that the galaxy population can be simply segregated into
star-forming galaxies and quiescent galaxies, and
assumes that redshift, $z$, and stellar mass, $M_*$, are the fundamental properties from which all of the other observables can be
statistically derived. In essence, drawing $z$ and $M_*$ from the observed stellar-mass functions for star-forming and quiescent galaxies, EGG uses
the SFR--$M_*$ `main sequence' displayed by star-forming galaxies to assign a SFR to each galaxy drawn for the star-forming mass function, and then utilises
empirical relations (established from the {\it HST}, {\it Spitzer} and {\it Herschel} observations of the CANDELS fields) to derive dust attenuation, optical colours and disk/bulge morphologies
for the galaxies. Dust attenuation is applied using the Calzetti law (\citealp{Calzetti:1994}; \citealp{Calzetti:2000})
while the redshift-dependent impact of the Inter-Galactic Medium (IGM) is calculated using the prescription of \cite{Madau:1995}.

After random scatter is introduced (to reflect the observed distributions of each parameter), \cite{Corentin:2017}
assign each galaxy a best-fitting SED and also calculate synthetic photometry by integrating the redshifted SED through any chosen set of filters as specified by the user.
Effectively, \cite{Corentin:2017} create a mock galaxy catalogue calibrated to the highest-quality existing data, with the ability to produce either SEDs or perfect (noise-free)
multi-wavelength photometry. 

For the current investigation we initially used EGG to create a sample of 3,908,096 galaxies across a sky area of 1.2\,deg$^2$, complete to a magnitude limit of 29.5\,mag in the
F440W NIRCam filter. The resulting synthetic galaxy population spans the redshift range $0 < z < 15$, and hence has a wide range of stellar masses ($5 < \log_{10}(M_*/{\rm M_{\odot}}) < 12$).
From this large-area sample we then constructed two galaxy sub-samples satisfying the area/depth of the two $\simeq 150-250$\,arcmin$^2$ regions produced by our observing strategy in the COSMOS and UDS CANDELS fields, which are the two most obvious equatorial extragalactic
survey fields for a major public NIRCam+MIRI survey, beyond CEERS and JADES (see section\, \ref{section4:Obs} and
Fig.\, \ref{Obs_Strat} for a more detailed description). The shapes of these two survey fields were defined using the existing {\it HST} mosaics, but
with realistic NIRCam+MIRI mosaicing strategies, taking into account
{\it JWST} roll-angle and minimum background constraints.

Having defined these two survey fields, we
adopted a straw-man approach to observing time, and calculated
what could be achieved if we committed $\simeq 100$\,hr of {\it JWST}
observing time to each field. The calculation of the
resulting noise levels for each
passband is described in detail in Section\,\ref{sec4:sim_obs}, but taking into account
the achievable sensitivities, we trimmed our EGG galaxy samples
by removing any non-detected objects (defined as $<4$-$\sigma$ detections in
more than 4 filters according to our proposed 8-filter NIRCam
observing strategy).

The final redshift and mass distributions for the predicted {\it JWST} COSMOS-CANDELS and UDS-CANDELS galaxy samples are shown in Fig.\,\ref{UDS_COS_Dist}.
The predicted number of detected galaxies is 69,766 and 51,879 in the UDS and COSMOS, respectively, and the sample properties are summarised in Table\, \ref{table:z_stat}.
The slightly different galaxy numbers and redshift/mass distributions shown
in Fig.\,\ref{UDS_COS_Dist} result from the necessarily slightly different observing strategies
adopted for each field (due to field geometry and allowed telescope roll angle); the proposed UDS survey is approximately 2$\times$ wider and therefore shallower than the COSMOS survey. Nevertheless, the predicted
galaxy populations are broadly very similar. The anticipated galaxy redshift distribution in both fields peaks at $z \simeq 0.5$, with median $z \simeq 2$, and extends to $z > 10$,
while the stellar-mass distribution is predicted to peak at $\log_{10}(M_*/{\rm M_{\odot}}) \simeq 8.5$ and extend to $\log_{10}(M_*/{\rm M_{\odot}}) > 11$. The UDS appears to yield far more lower-redshift galaxies than our deeper COSMOS observations, yet a similar number of high-redshift objects are detected in both fields. This is largely caused by the increased detection of higher-mass galaxies found at lower-redshift, while the COSMOS imaging is better suited to detecting the higher-redshift objects and/or lower-mass objects in our deep field.

In the next section we explain how we added noise to the photometry of the simulated galaxies, and summarize the realistically achievable depths given our adopted
strawman observing strategy.

\section{Simulating Observations}\label{sec4:sim_obs}
In this section we describe the processes by which we simulated the {\it JWST} NIRCam and MIRI data for our model galaxies, and then outline the strawman observing strategy 
which leads on to the typical depths anticipated in each NIRCam and MIRI filter.

\subsection{NIRCam Simulations}\label{section4:Pandeia}

The NIRCam observations were simulated using the simulation-hybrid {\sc python} engine, Pandeia \citep{Pandeia:2017}. Pandeia calculates 2-D pixel-by-pixel signal-to-noise ratios for all observing modes onboard  {\it JWST}.  The {\it JWST} Exposure Time Calculator (ETC) web interface is based on this same {\sc python} engine. However, for large numbers of objects and simulations, the web interface is not sufficient. Pandeia allows the user to define an astrophysical scene with an external input SED, and then to `observe' the synthesized objects using observational parameters defined by the user. The input SEDs can be normalized at any wavelength, across all {\it JWST} and {\it HST} filters. 

The astronomical scene comprises a set of sources and associated SEDs independent of the observing instrument. This scene is projected onto a realised scene, that is initially spatially sub-sampled, avoiding under-sampling of the PSF. The realized scene cube is then projected onto an appropriate detector plane, dictated by the chosen instrument and the observing mode defined by the user. The details of how this is done are described in \cite{Pandeia:2017}. 

In our case, we define our observing strategy using background models, an instrument mode and a specified detector setup. The background models are chosen based on the region of the sky where JWST is pointing at a user-defined time of year. The detector setup governs the specified read-out mode along with the number of groups per integration, number of integrations per exposure and the total number of exposures. Combing these parameters provides a total exposure time for our observations in each selected filter. These values were selected based on the average exposure times calculated from our APT-designed observing strategy described in section\, \ref{section4:Obs}. We note that we assume the objects are point sources.

The intermediate step of tracking the photons through the optical train to the detector is done in a similar manner to our independent MIRI simulator (described next in section \ref{MIRISIM}). Once the SNR has been found in each photometric band for our chosen integration times, we assign an error and scatter the data accordingly (assuming a Gaussian distribution). The result is a best estimate of the anticipated NIRCam photometry (and uncertainty) for each of our simulated galaxies, after processing with the NIRCam data-reduction pipeline.

\subsection{Independent MIRI Simulator}\label{MIRISIM}
We calculated the MIRI photometry using an independent code which we had already created prior to this work based on a set of sensitivity models used in the official MIRI Team simulator, MIRISim \citep{Glasse:2015}. 
We could of course have used the Pandeia software to perform SNR calculations in the MIRI bands, but it transpires to be far less efficient than our own simulator; for a few objects this would not be important, however, as we wished to run more than 20,000 galaxies through our pipeline under different observing scenarios, Pandeia can become very computationally expensive. 
Our own MIRI simulator quickly produces very accurate synthetic observations to within a $\sim$1\% accuracy of that predicted by the official ETC for all observing modes with MIRI. We show the results of a simple test run to asses the quality and accuracy with which our own independent simulator compares to that of the ETC in the appendix (Fig.\,\ref{fig:sim_test}).  After calculating the expected photometric errors for our chosen observing configuration, we again scattered the predicted photometry using a Gaussian distribution to provide the raw observational data expected from our MIRI imaging of the simulated (BC03 and EGG) galaxies. Full details of our MIRI simulator are published in Kemp et al. (in preparation), but for the interested reader we provide a brief summary
of its key elements (and the underlying physics) in Appendix A.

\subsection{Observing Strategies}\label{section4:Obs}

\begin{figure}

\centering
   \includegraphics[scale=0.35]{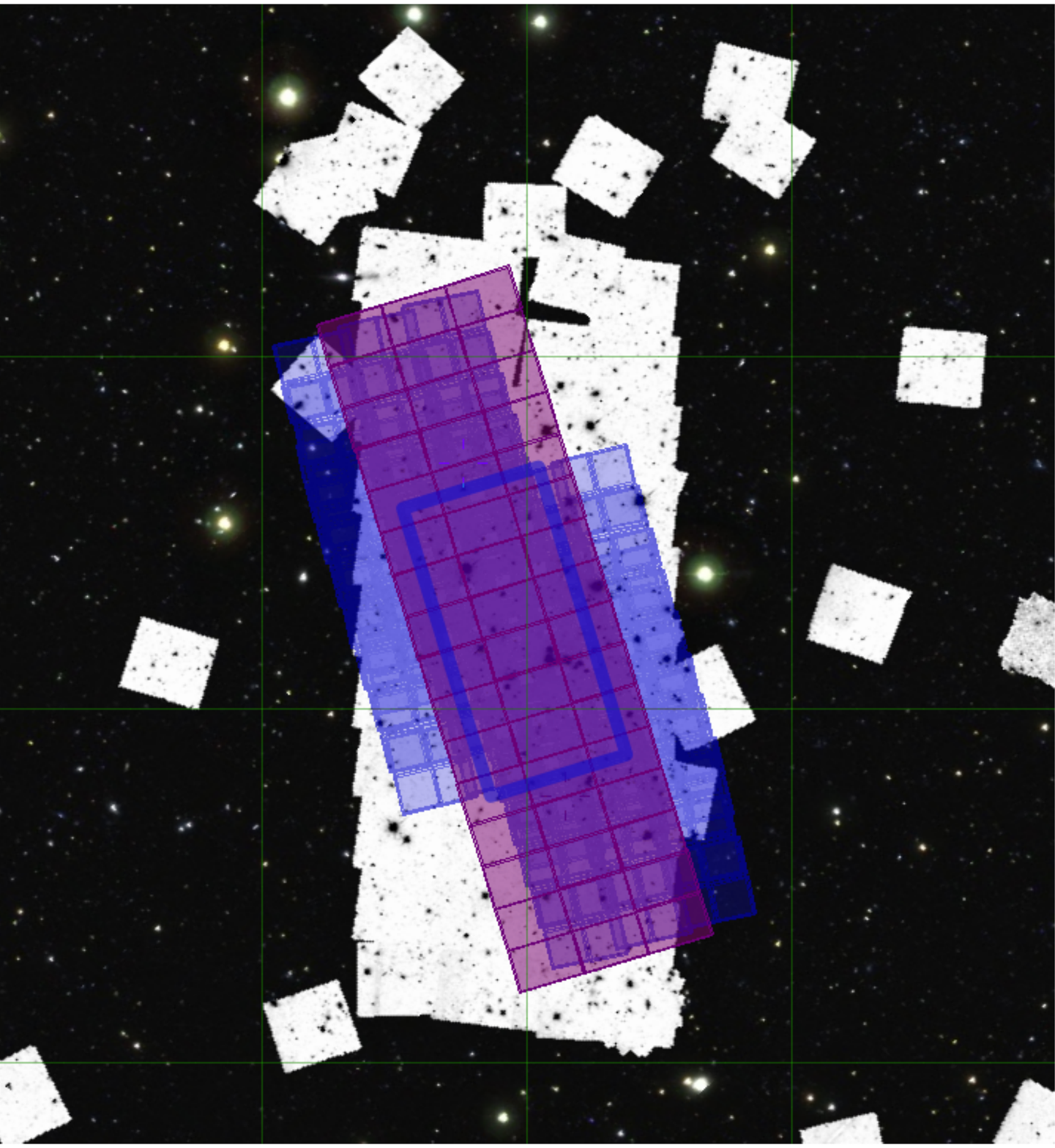}
   \caption{The layout for our prescribed COSMOS observing strategy described in section\, \ref{section4:Obs}. The red tiles represent the dithered MIRI coverage in the F770W filter, while the blue tiles represent the dithered NIRCam pointings with the deepest region (due to multiple exposures) highlighted by the dark blue box. We overlay this mosaicing strategy onto the current HST ancillary data available in COSMOS CANDELS along with the UVISTA colour composite image.}
         \label{Obs_Strat}
\end{figure}

Clearly the final observing strategy of any large-scale public extragalactic survey with {\it JWST} cannot be known until {\it JWST} is operational, and open-time programmes are selected for the early
observing cycles. However, our aim here is to provide predictions which are both generic and realistic. We have therefore explored in detail possible NIRCam+MIRI imaging strategies for the 
UDS-CANDELS and COSMOS-CANDELS covering the survey areas outlined in Table\, \ref{table:z_stat}.  As a result we have produced a fully-realistic, schedulable,`strawman' observing strategy which devotes $\simeq 100$\,hr of {\it JWST} observing time to each of these two fields, while striving to maintain $> 50$\% efficiency: specifically our proposed observing scenario devotes 101.9\,hr of observing yielding 53.4\,hr  of science time in the UDS-CANDELS field, and 90.6\,hr of observing time yielding 47.5\,hr of science time in the COSMOS-CANDELS field. The mosaic footprint for our COSMOS CANDELS field imaging strategy is shown in Fig.\, \ref{Obs_Strat}. We overlay the proposed observing strategy for our MIRI (red tiles) and NIRCam (blue tiles) pointings on the current HST and UVISTA COSMOS coverage. The UDS observing strategy follows a different format due to more difficult V3PA constraints (perpendicular to COSMOS). Both the UDS and COSMOS strategies were built using the APT to ensure a fully realistic and schedulable set of observations.

As explained above, the proposed observing plan
involves equal integration times in 8 NIRCam filters (F090W, F115W, F150W, F200W, F277W, F356W, F410M, F444W), with MIRI observations obtained in parallel over as much of the NIRCam 
area as possible given the constraints of survey field size/shape and the fairly limited permitted {\it JWST} roll angles for equatorial target fields. For this strawman survey we devote all of the parallel 
MIRI integration time to F770W imaging since, as already explained, the focus of this paper is on the value of the added 7.7\,$\mu$m MIRI photometry to the NIRCam 
photometric dataset. The anticipated 5-$\sigma$ point-source depths (for all 10 NIRCam+MIRI filters) predicted by our noise simulations for the deepest region of contiguous area coverage in each field under our prescribed observing strategy are summarized 
in Table\, \ref{tbl:filter_depths}. The typical depths are somewhat lower in the UDS-CANDELS field because, as already explained above, the field geometry leads naturally to a somewhat wider-area, shallower survey
(assuming comparable intergation time in each field) if substantial MIRI+NIRCam overlap is to be achieved. We stress, however, that the main results of this work, described in the next section,
apply to any similar NIRCam multi-band survey with MIRI parallel observations, and in that sense our conclusions can be regarded as generic, even if based on a specific strawman 
observing strategy to ensure realism.

Given these depths (summarized in Table\,\ref{tbl:filter_depths}), as previously mentioned in relation to Table\,\ref{table:z_stat} and Fig.\,\ref{UDS_COS_Dist}, in the interests of efficiency and meaningful analysis,
we removed any galaxies from the simulations that did not match certain criteria as judged from the simulated photometry. Specifically `non-detections' were removed, as judged by synthesised photometry which yielded less than 4-$\sigma$ detections in more than four {\it JWST} wavebands. 

Finally, we assumed that we can utilise ancillary homogeneous {\it HST} optical imaging coverage of our survey fields at the depths/filters of the current public ACS data. For the
UDS-CANDELS and COSMOS-CANDELS fields, this consists of $V_{606}$ and $i_{814}$ imaging which, in both bands and in both fields reaches a 5-$\sigma$ point-source detection limit of
28.4\,mag. As with the {\it JWST} photometry, we simulated and scattered the predicted photometry in the {\it HST}/ACS bands given the known depths, and assuming Gaussian errors.

\section{Reclaiming Galaxy Properties}\label{sec5:reclaim}

In this section we introduce the SED-fitting code \textsc{Bagpipes} and explain the process of reclaiming the physical properties of the simulated galaxies 
from the synthesized {\it JWST} photometry. We also present and analyse the results for our BC03 and EGG simulations, exploring the impact of the addition of both MIRI and {\it HST} photometry
to the core NIRCam 8-filter data.

\subsection{SED Fitting Code - \textsc{Bagpipes}}\label{section5:Bagpipes}

Bayesian Analysis of Galaxies for Physical Inference and Parameter Estimation (\textsc{Bagpipes}, \citealp{Carnall:2018}) is a Bayesian spectral-fitting code, capable of fitting galaxy SEDs
to photometric and/or spectroscopic data simultaneously in a statistically robust manner. \textsc{Bagpipes} can also be used to also generate model galaxy spectra based on set physical parameters described via {\sc python} dictionaries; all of our simulated galaxy SEDs based on the Bruzual \& Charlot (2003) spectrophotometric models were generated using \textsc{Bagpipes}, with the specified physical parameters summarized in Tables 1 and 2.

\textsc{Bagpipes} requires defined multiple-model parameterisation to fit against the submitted observational data, including the uncertainties of the observations. The user is required to define a prior probability distribution, describing the free parameters and limits for the fitting process. For this work we varied galaxy redshift ($0 < z < 15$), stellar mass ($0 < \log(M_*/{\rm M_{\odot}})
< 13$), metallicity ($0 < (Z/{\rm Z_{\odot}}) < 3$) and dust obscuration ($0 < (A_V/{\rm mag}) < 7$). All galaxies are fitted against two separate SFHs, exponentially decreasing/increasing and Log-Normal (see Section 2.3 in \citealp{Carnall_2:2018}) . We allow the exponential prescription to vary in both decay timescale ($-10 < \tau < 10$) and age ($0.001 < T/{\rm Gyr}) < 10$), thus allowing SFHs to be exponentially increasing or decreasing, and spanning a very large region of parameter space. For the Log-Normal SFH, we allow a wide range in Full Width Half-Maximum ($0 < (FWHM/{\rm Gyr}) < 10$) and maximum age 
($0.001 < (t_{max}/{\rm Gyr}) < 10$). Through attempting to fit this range of models to the simulated photometry, \textsc{Bagpipes} produces a posterior distribution for each free parameter, and hence calculates the posterior median along with the $\pm$1-$\sigma$ confidence intervals for each parameter value. The free parameters and their allowed ranges are summarized in Table\, \ref{table:fit_params}.

\begin{table}
\caption {The free parameters, and their allowed ranges, used to build the SED library within \textsc{Bagpipes} for fitting to our simulated galaxy photometry.}\label{table:fit_params}
  \begin{center}
    \begin{tabular} { c | c | c }
      \hline
      \textbf{Parameter} & \textbf{Description}  & \textbf{Range}  \\
      \hline
            $M_*$  & Stellar Mass & $0 < \log_{10}(M_*/{\rm M_{\odot}}) < 13$ \\
          $Z$ & Metallicity & $0 < (Z/{\rm Z_{\odot}}) < 3$ \\
          $z$ & Redshift  & $0 < z < 15$\\ 
          $A_V$ & Dust Obscuration & $0 < (A_V/{\rm mag}) < 7$ \\
          \hline
          & \textbf{Exponential SFH} & \\
          \hline
          $\tau$ & Timescale & $-10 < (\tau/{\rm Gyr}) < 10$   \\
          $T$ & Age of SFH & $0.001 < (T/{\rm Gyr}) < 15$  \\
          \hline
          & \textbf{Log-Normal SFH} & \\
          \hline
          $FWHM$ & SFH width & $0 < (FWHM/{\rm Gyr}) < 10$   \\
          $t_{max}$ & Age of SFH & $0.001 < (t_{max}/{\rm Gyr}) < 15$  \\
    \end{tabular}
  \end{center}
 
\end{table}  

\subsection{Results for the BC03 simulated galaxy sample}

        \begin{figure*}
   \centering
   \includegraphics[width=15cm]{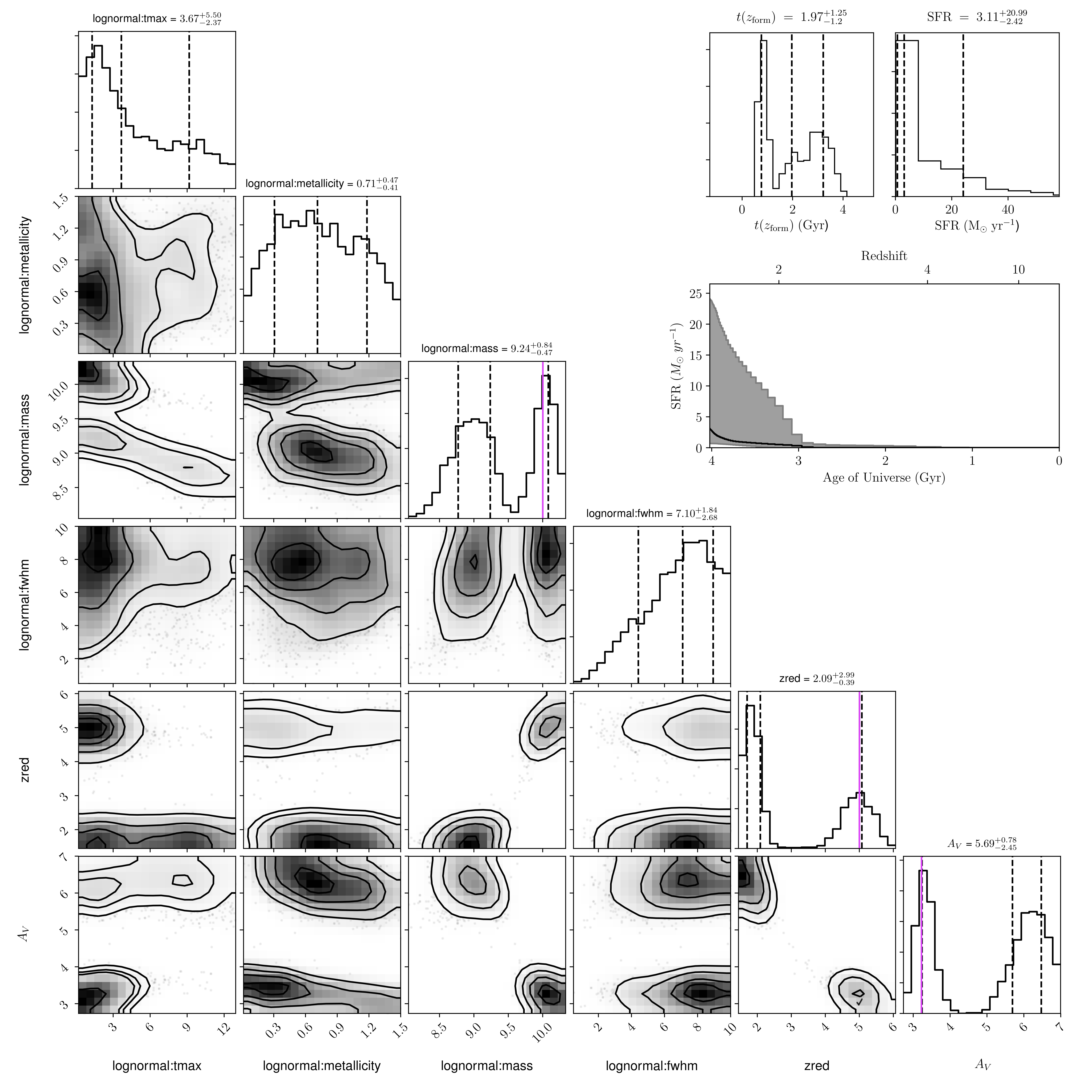}
   \includegraphics[width=15cm]{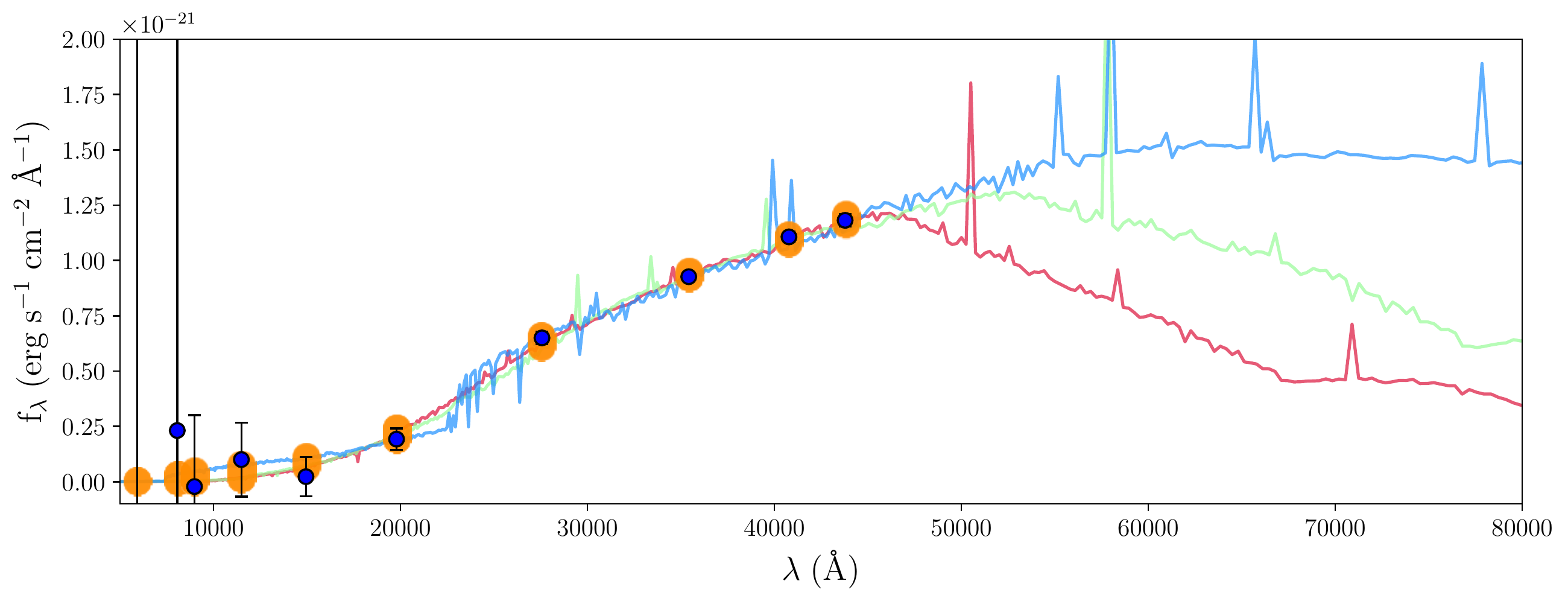}
      \caption{An example of a \textsc{Bagpipes} fit to a dusty ($A_V = 3.2$\,mag), $z = 5$ galaxy without any MIRI photometry (i.e. fitting only to the photometry from the 8 NIRCam filters). 
The upper panel presents a corner-plot of the resulting posterior distributions of all 6 free parameters described in Table\,\ref{table:fit_params}. We also show the inferred age, SFR and SFH. The dashed lines in each panel represent the posterior median along with the $\pm$1-$\sigma$ solutions. The solid purple line marks the true input values for the stellar mass, redshift and dust obscuration. The lower panel shows our simulated galaxy photometry (blue points with error-bars), along with the {\sc Bagpipes} SED fits to these data. The green line represents the SED extracted from the posterior medians of each distribution, in this case, at redshift $z \simeq 2.09$. The red and blue lines show the corresponding $-$1-$\sigma$ and +1-$\sigma$ SEDs respectively. In this case, the blue line corresponds to a high-redshift solution (at $z \simeq 5$, close to the true value), while the red line corresponds to a low-redshift solution (at $z \simeq 1.8$, close to the stronger low-redshift peak seen in the redshift corner plot). Most parameter distributions appear to be dual-peaked, with a larger proportion of the redshift distribution lying in the low redshift solution ($z \simeq 2$, red SED).}\label{fig:eg_n7}
   \end{figure*}

\begin{figure*}
   \centering
   \includegraphics[width=15cm]{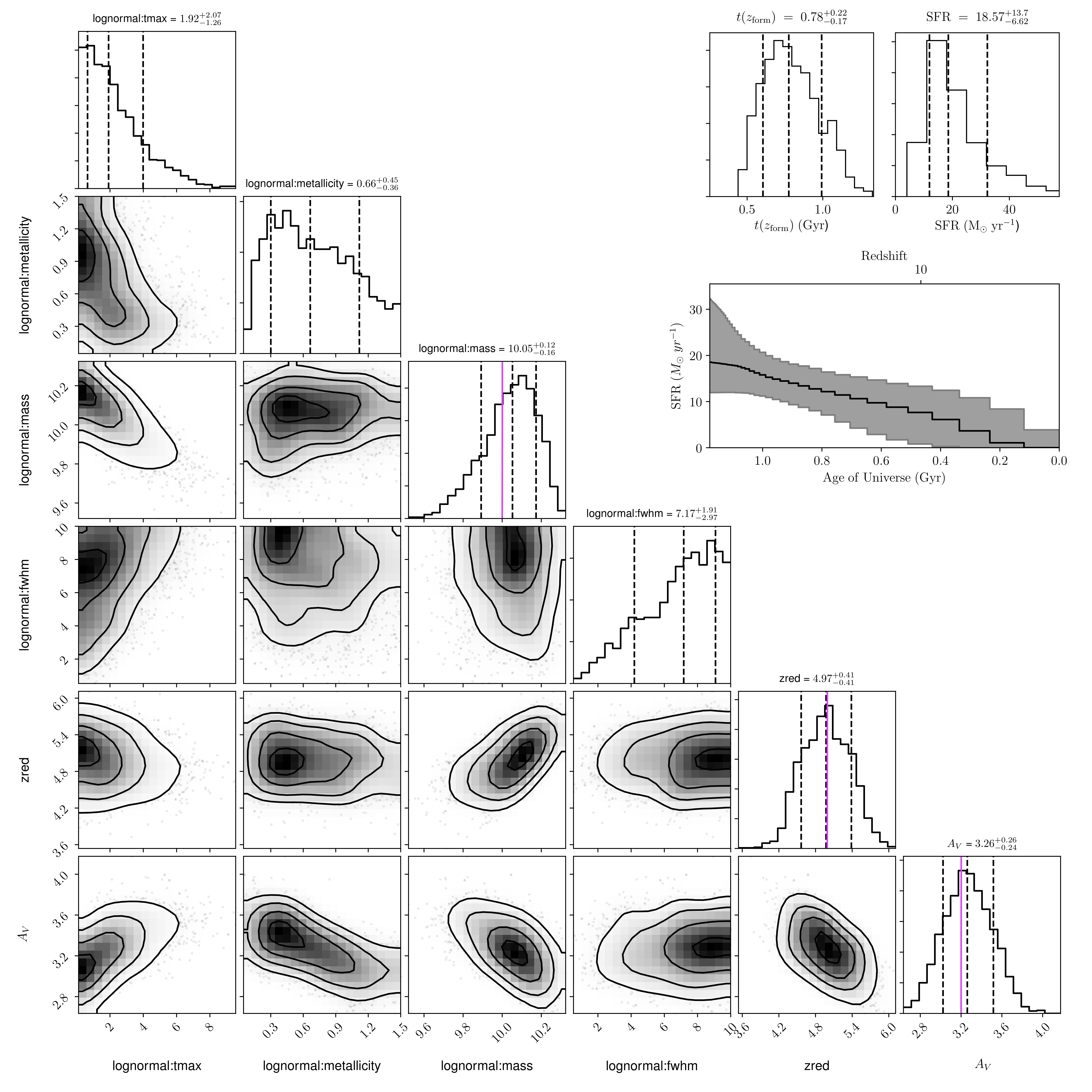}
    \includegraphics[width=15cm]{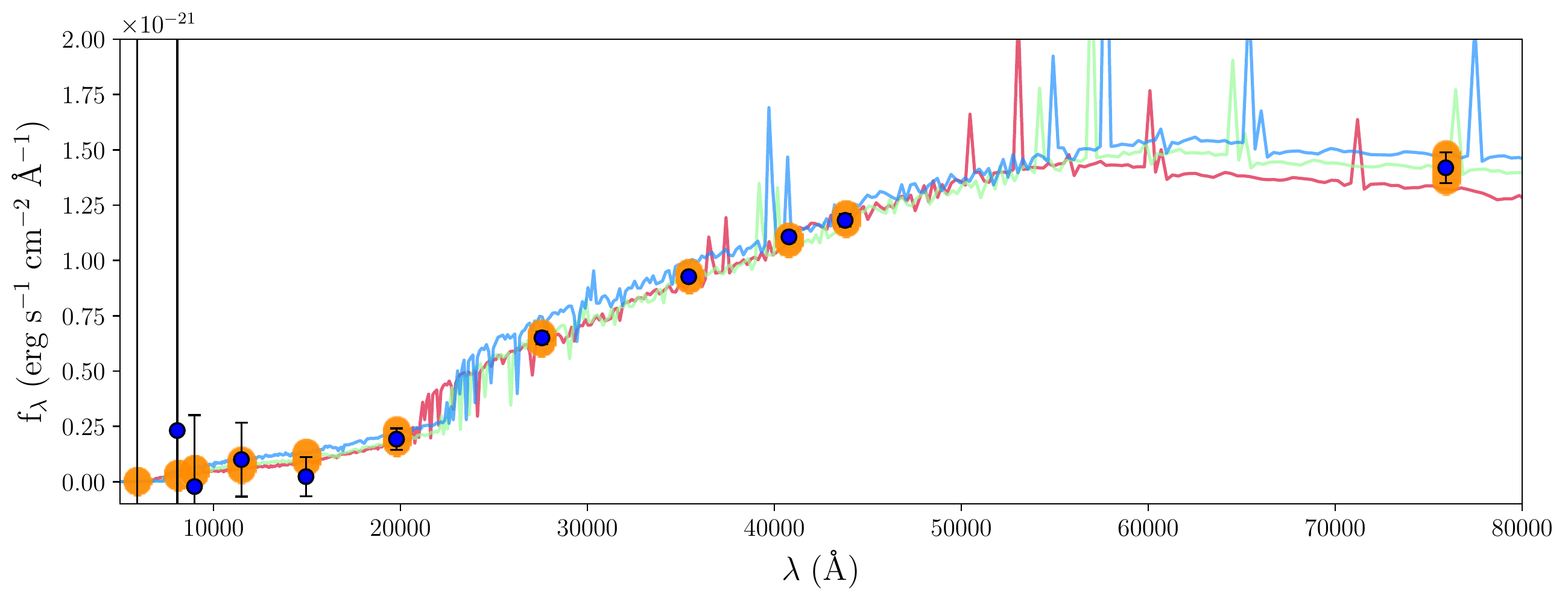}
      \caption{The \textsc{Bagpipes} fit to the same $z = 5$ galaxy as in Fig.\,\ref{fig:eg_n7}, but this time also fitting to the simulated MIRI F770W photometry. Again, the upper panel presents a corner-plot of the resulting posterior distributions of all 6 free parameters described in Table\, \ref{table:fit_params}, and we also show the inferred age, SFR and SFH. As before, the dashed lines in each panel  represent the posterior median along with the $\pm$1-$\sigma$ solutions, but this time it can be seen that the parameter distributions are no longer dual-peaked and sit close to the true input values represented by the solid purple line. The lower panel shows our simulated galaxy photometry (blue points with error-bars, this time including the MIRI 7.7-$\mu$m photometry), along with the {\sc Bagpipes} SED fits to these data. The green line represents the SED extracted from the posterior medians of each distribution, this time at redshift $z \simeq 4.97$ (only $\Delta z = 0.03$ from the true value). The red and blue lines show the corresponding $-$1-$\sigma$ and +1-$\sigma$ SEDs respectively; these now quantify the width of the single peak close to $z = 5$ ($z = 4.56$ and $z = 5.38$) due to the fact that the addition of the MIRI photometry has removed the bimodality in the distributions.} 
        \label{fig:eg}
\end{figure*}

In Fig.\,\ref{fig:eg_n7} and Fig.\,\ref{fig:eg} we show an example of using {\sc Bagpipes} to fit the simulated photometry of a 
 $z$ = 5, $M_* = 10^{10}\,{\rm M_{\odot}}$, dusty ($A_V$ = 3.2\,mag)  galaxy, fitting to {\it HST}+NIRCam and then {\it HST}+NIRCam+MIRI F770W photometry respectively.
In each case, we show the posterior distributions and contour plots for all the fitted free parameters described in Table\,\ref{table:fit_params}, including also the inferred SFH (top right). We present the median and $\pm$1-$\sigma$ best-fit SEDs to our scattered photometric data in the lower panel of each figure. 

In the first example (Fig.\,\ref{fig:eg_n7}),  the redshift posterior distributions and contours plots are dual-peaked, with the majority of the distribution lying within the low-redshift 
peak at $z \simeq 2$. As a result, the posterior median (middle dashed line) lies within the incorrect, low-redshift peak of the solution. A separate peak in the redshift distribution
can be seen close to the correct (simulated) redshift at $z \simeq 5$, however the dominance of the low-redshift solution means that this is only just included within the $+$1$\sigma$ confidence region.

Incorrectly inferring the redshift has an obvious detrimental impact on the estimation of galaxy mass, especially given this type of bi-modal posterior 
distribution. The stellar-mass distribution and contour plots also therefore show two potential solutions, peaking at approximately $M_* \simeq 10^{9}\,{\rm M_{\odot}}$ and
$M_* \simeq 10^{10}\,{\rm M_{\odot}}$. The width of the distribution lying around the incorrect (lower) stellar-mass estimate at
$M_* \simeq 10^{9}\,{\rm M_{\odot}}$, draws the posterior median to lie at approximately $10^{9.25}\,{\rm M_{\odot}}$.
This type of serious mass under-estimate (clearly linked to the bi-modal distribution of redshift) can introduce substantial errors in the determination 
of the galaxy stellar-mass function, and its inferred evolution with cosmic time.

The best-fitting SEDs are shown alongside the galaxy photometry in the bottom panel of Fig.\,\ref{fig:eg_n7}. The blue points represent the raw photometry (with black error bars),
while the orange `points' show the photometric posterior distribution. Three SEDs are shown, with the red, green and blue SEDs showing the best fitting SEDs for the
posterior median (green) and $\pm$1-$\sigma$ (blue and red, respectively). These can be matched to the dashed lines on the posterior distributions found in the corner plots:
the blue line is a  $z \simeq 5$, $M_*~\simeq~10^{10}\,{\rm M_{\odot}}$, $A_V = 3.2$\,mag galaxy, while the green line represents the median $z = 2.09$, $M_* = 10^{9.25}\,{\rm M_{\odot}}$, $A_V = 5.7$\,mag SED fit,
and the red line is a $z = 1.7$, $M_* = 10^{8.77}\,{\rm M_{\odot}}$, $A_V =6.5$\,mag galaxy.

\begin{figure*}
\centering
\includegraphics[width=15cm]{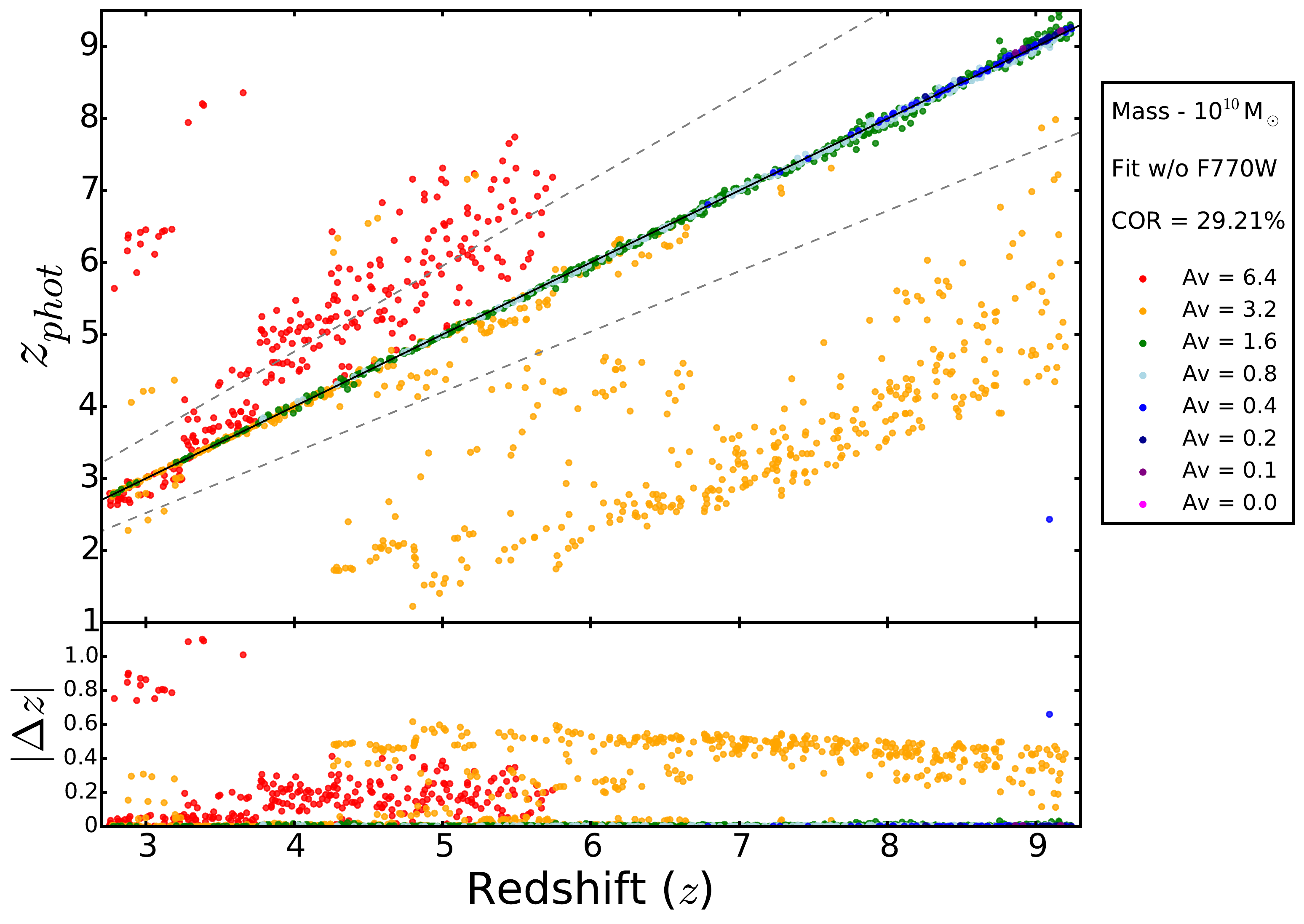}
\includegraphics[width=15cm]{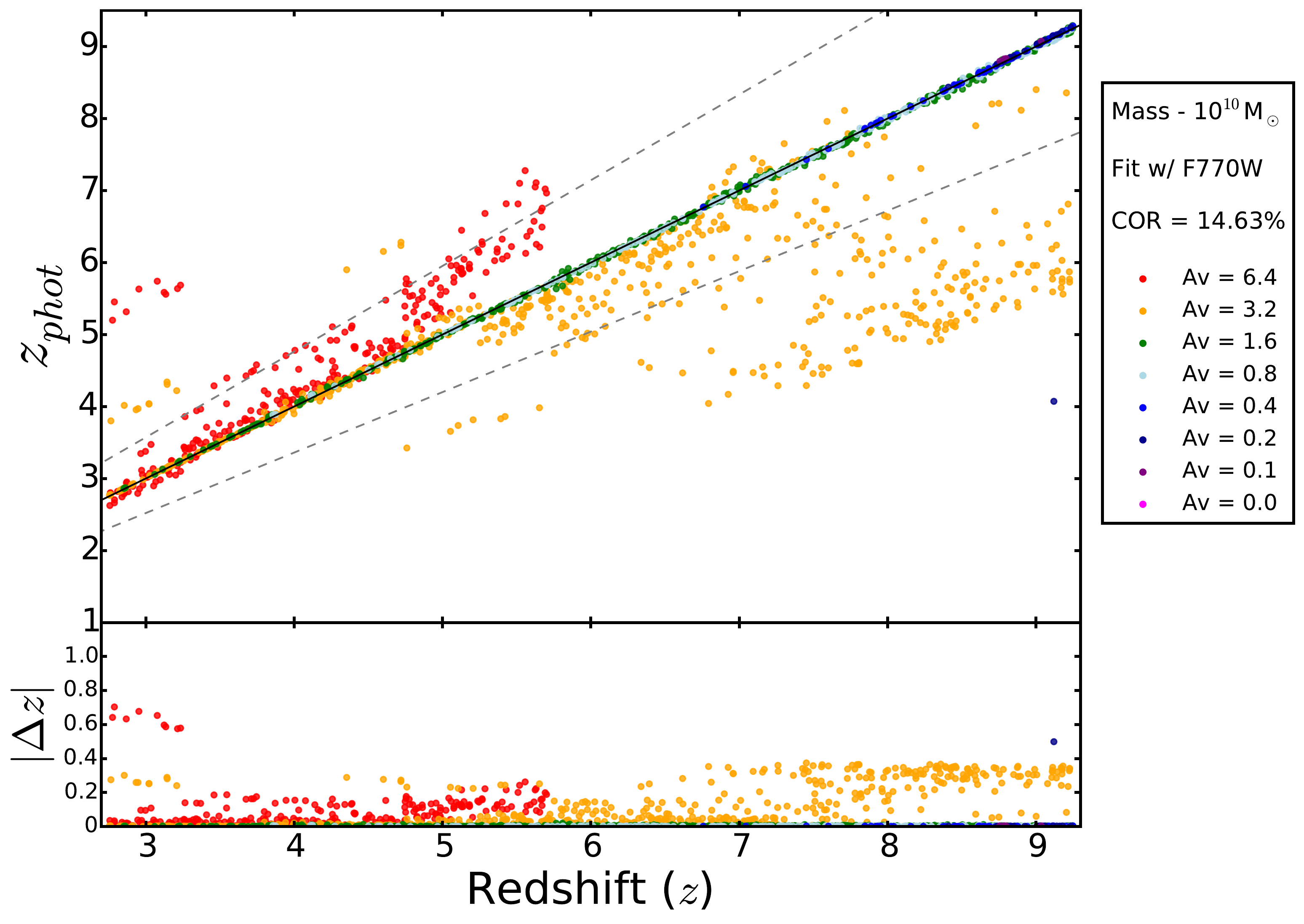}
\caption{Photometric versus input redshifts (with $|\Delta z|$ = $|(z_{phot} - z_{input})/(1+z_{input})|$ sub-plotted below each panel) 
for the $10^{10}\,{\rm M_{\odot}}$ 
exponentially-increasing SFH BC03 simulated galaxies. As indicated in the legend, the data points are coloured according to 8 bins in simulated 
dust obscuration ($A_V = $0, 0.1, 0.2, 0.4, 0.8, 1.6, 3.2, 6.4 mag). The upper panel shows the results when the galaxies are fitted without including the MIRI F770W photometry 
in the simulated dataset, while the lower panel shows the impact of including the MIRI photometry. The legend on the right-hand side of both panels 
gives the catastrophic outlier rates (COR), where the COR is defined by the percentage of galaxies for which  $|\Delta z| > 0.15$. It can be seen that the photometric redshifts 
are of excellent quality for all but the dustiest galaxies ($A_V > 3$\,mag) and that the addition of the MIRI data improves the results for a significant subset of these, particularly 
at intermediate redshifts (as illustrated by the example shown in Fig.\,\ref{fig:eg_n7} and Fig.\,\ref{fig:eg}).}\label{Fig:M10_zz}
\end{figure*} 

In Fig.\,\ref{fig:eg} we show the results of fitting the same simulated galaxy, but this time including MIRI F770W photometry in the simulated dataset.
It can be seen that the posterior distributions are very different, and in particular no longer dual-peaked. The distributions for redshift,
stellar mass and dust obscuration all take the form of a  single sharp peak lying very close to the correct values. This results in much tighter $\pm$1-$\sigma$ confidence intervals
with a relatively narrow range of allowed parameter values. We also reclaim the shape of the SFH relatively well, with the fitting producing a SFR that grows slowly with time,
matching very well the input fiducial $\tau = -0.1$ SFH shown in Fig.\,\ref{figure:tau_fit}. 

The scatter seen in the fitted SEDs also reduces dramatically with the addition of the F770W filter. This time the 
SED extracted from the posterior medians (green line) lies at redshift $z \simeq 4.97$ (only $\Delta z = 0.03$ from the true value), and the
red and blue lines (showing the corresponding $-$1-$\sigma$ and +1-$\sigma$ SEDs respectively) now simply quantify the width of the single peak close to $z = 5$
($z = 4.56$ and $z = 5.38$) due to the fact that the addition of the MIRI photometry has removed the bimodality in the distributions.

For this type of dusty, red, high-redshift galaxy the impact of adding MIRI photometry is thus clear. It is also clear that the {\it HST} ACS optical data (in this case $V_{606}$ and $i_{814}$)
has very little impact: the red colour of this galaxy combined with the depth of available ACS imaging in the CANDELS fields means that the errors on the ACS photometry are too large
(in comparison with the {\it JWST} photometry) to significantly affect the fit. 

\begin{figure*}
   \centering
   \includegraphics[width=18cm]{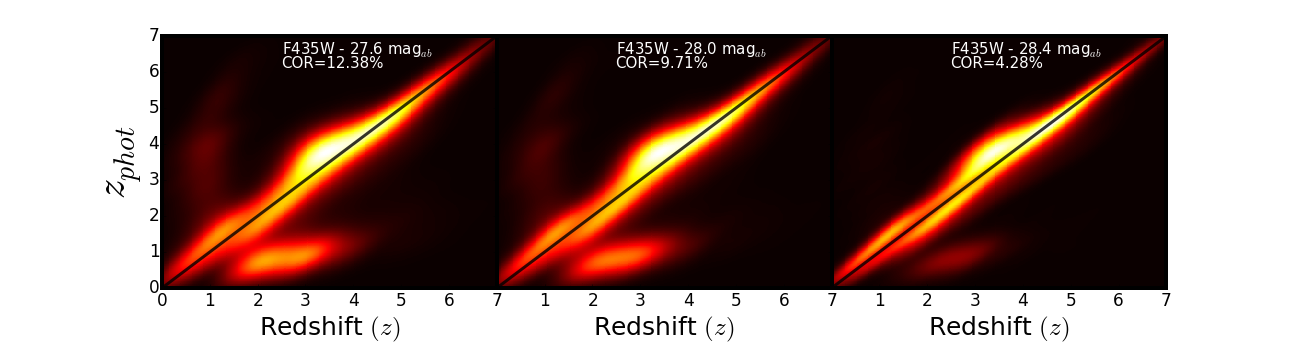}
   \caption{Density maps showing reclaimed photometric redshift as a function of simulated galaxy redshift for our sample of EGG galaxies,
     exploring the impact of increasingly deep {\it HST} ACS $B_{435}$ imaging as indicated in each panel.
     Left: Photometric redshifts reclaimed using the 8 NIRCam Bands with existing {\it HST} $V_{606}$, $i_{814}$ and $B_{435}$ imaging (reaching 27.6\,mag, 5-$\sigma$).
     Centre: The results of the same procedure, but assuming the $B_{435}$ imaging is 0.4\,mag deeper. Right: The results when it is assumed that the $B_{435}$ imaging is 0.4\,mag deeper still.
     The catastrophic outlier rate (COR) in each case is also given, defined as the percentage of galaxies with $|\Delta z|$ $>$ 0.15, where
     $|\Delta z|$ = $|(z_{phot} - z_{input})/(1+z_{input})|$.}\label{Fig:EGG_zz}
\end{figure*} 

\begin{figure*}
   \centering
      \includegraphics[width=18cm]{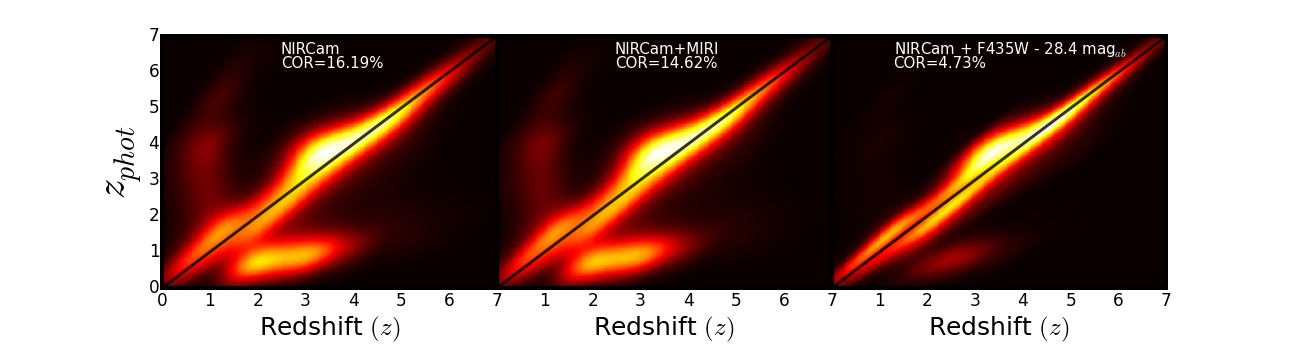}
                \caption{Density maps showing reclaimed photometric redshift as a function of simulated galaxy redshift for our sample of EGG galaxies in multiple filter combinations.
                  Left: Photometric redshifts reclaimed using the 8 NIRcam Bands with existing {\it HST} $V_{606}$ and $i_{814}$ imaging. Centre: Photometric redshifts reclaimed using the 8 NIRcam Bands
                  with existing {\it HST} $V_{606}$ and $i_{814}$ + MIRI F770W photometry.  Right: Photometric redshifts reclaimed using the 8 NIRcam Bands with
                  {\it HST} $V_{606}$ and $i_{814}$ + $B_{435}$ (28.4\,mag) imaging.
                  The catastrophic outlier rate (COR) in each case is also given, defined as the percentage of galaxies with $|\Delta z|$ $>$ 0.15, where
                  $|\Delta z|$ = $|(z_{phot} - z_{input})/(1+z_{input})|$.}\label{Fig:EGG_zz_2}
\end{figure*} 

Having highlighted an example of an object for which MIRI photometry clearly matters, in Fig.\,\ref{Fig:M10_zz},
we plot derived photometric redshift versus true input galaxy redshift for our sample of $M_* \simeq 10^{10}\,{\rm M_{\odot}}$ BC03-simulated galaxies over the redshift range $3 < z < 9$,
in this case (for clarity) focussing on
the subset of such galaxies simulated with exponentially-rising SFHs. These plots show the results of attempting to determine photometric
redshifts without (upper panel) and then with (lower panel) inclusion of the MIRI data. Both the $z-z$ plot and the normalised redshift distribution ($|\Delta z|$ = $|(z_{phot} - z_{input})/(1+z_{input})|$)
are plotted in each case, and the galaxies are colour-coded by (input) dust obscuration. It is important to note that although we have 13 discrete redshift bins, we scatter the true input redshift values on the x-axis randomly by up to $\pm$$0.25 \delta z$. We subsequently apply the same offset to the fitted photometric redshifts on the y-axis to effectively project the redshift onto smaller, equally sized redshift bins making the plot itself far more digestable. Fig.\,\ref{Fig:M10_zz} confirms that the galaxy discussed above (Fig.\,\ref{fig:eg} and Fig.\,\ref{fig:eg_n7}) is indeed typical
of the sort of object for which the addition of MIRI 7.7-$\mu$m photometry is important. The addition of the MIRI data has no real impact on the (already excellent) photometric
redshifts for the vast majority of objects with $A_V < 2$\,mag, but clearly impacts on the (often much poorer) results for the galaxies in the two highest bins of obscuration (essentially
$A_V > 2$\,mag). The catastrophic outlier rate (COR, defined as $|\Delta z|$ $>$ 0.15) is reduced by a factor of two (from COR = 29.2\%\ to COR = 14.6\%)
by the inclusion of the MIRI photometry, with the improvement most dramatic at intermediate redshifts ($z \simeq 4 - 6$). At very high redshifts the improvement is much less dramatic, primarily
because dusty galaxies of this mass yield relatively poor-quality detections at $z > 7$ in the F770W filter (given our chosen exposure times).
Finally, we note here that while the results shown in Fig.\,\ref{Fig:M10_zz} include {\it HST} ACS photometry, the effect of excluding any {\it HST} ancillary data only degrades the COR by a very small amount
($\simeq 2$\%) but, as we discuss further in the next section, this is largely because {\bf i)}~the impact of {\it HST} optical data is most important at redshifts $z < 3$, and {\bf ii)} the BC03 simulated
galaxy sample analysed here contains a (deliberately) unrealistically-large proportion of dusty galaxies with extreme values of $A_V$. The approach taken here has therefore
enabled us to isolate the {\it type of galaxy} for which MIRI parallel-mode imaging is of real value in contraining model parameters.

How important such dusty high-redshift galaxies actually are obviously
depends on their prevalence/distribution in the general galaxy population, and the science question to be addressed. We note that, while we have assumed that such objects exist in all three mass bins of our BC03 simulated galaxy sample ($M_*~=~10^{10}, 10^{9}, 10^8\,{\rm M_{\odot}}$), it transpires that, in the two lower mass bins, very few such objects survive our S/N cuts due to the combination of low mass, high
redshift, and extreme dust attenuation. Thus, even if such objects exist at such masses, they do not feature in our simulated {\it JWST} samples.
In summary, given realistic exposure times, it is only at stellar masses $M_* \ge 10^{10}\,{\rm M_{\odot}}$ that MIRI data can have a significant impact on the accuracy with which the physical properties
of dusty high-redshift galaxies are reclaimed. However, there is in any case a growing body of evidence that dust content is strongly correlated with galaxy stellar mass, and hence that such objects
are indeed largely confined to the stellar mass range  $M_* \ge 10^{10}\,{\rm M_{\odot}}$ in the real Universe (\citealp{Dunlop:2017}; \citealp{McLure_2:2018}).

\subsection{Results for the EGG simulated galaxy sample}

Finally, we used {\sc Bagpipes} to reclaim the properties of the galaxies in the simulated EGG galaxy samples in the COSMOS and UDS simulated surveys as summarized in Table\,\ref{table:z_stat}. 
Again we explored the impact of adding the MIRI photometry to the {\it HST}-ACS + {\it JWST}-NIRCam dataset, and this time found that the addition of parallel MIRI F770W
photometry had very little impact on the accuracy of the reclaimed photometric redshift estimation. In fact, the photometric redshifts improved
for only $\simeq$1\% of the EGG-simulated {\it JWST}-detected galaxy population, which is of course intended to produce a more realistic distribution of galaxy properties than explored in the BC03
simulations. There are two reasons for the relatively low impact of MIRI. First, the EGG simulations extend down to essentially $z \simeq 0$, and the redshift distributions are dominated by galaxies at
$z < 3$. Such objects, which were deliberately not included in the BC03 simulations (which were aimed at exploring the high-z galaxy population) have redshifts that are already well-constrained
by multiple detections in the ACS+NIRCam filters, or if not are too faint for MIRI to help. Second, binning the EGG sample by dust obscuration,
we found that the EGG simulations contained only a tiny number of dusty galaxies
with $A_V > 3.0$\,mag in the simulated {\it JWST} survey galaxy samples.
This may reflect a problem with EGG which requires further exploration/refinement, since such objects are certainly known to exist in the CANDELS fields (\citealp{Dunlop:2007}),
but at least it explains the virtually negligible impact of MIRI on the photometric redshifts (and hence stellar masses etc).

In fact, we found that where MIRI was making its very modest impact on the EGG photometric redshifts was for a small sample of relatively dust-free galaxies which happened to have
two alternative peaks in their posterior redshift distributions, and the addition of the MIRI F770W photometry just happened to tip the probability balance slightly in favour of the correct
peak. However, for these objects, the impact of MIRI was still too modest to modify the posterior distributions to the extent that the incorrect solutions could be
excluded, and on close examination of this subset of objects we discovered that the real problem was the lack of {\it HST} data at wavelengths shortward of
the existing $V_{606}$ data. We therefore decided to explore whether the addition of ACS $B_{435}$ imaging in these fields would actually
have more impact on this subset of objects than the addition of the MIRI photometry. To explore the impact of adding $B_{435}$ imaging of increasing depth, 
we decided to perform a series of re-runs using differing $B_{435}$ depths of 27.6, 28.0 and 28.4\,mag. The first of these depths was chosen to represent what could be achieved
in a Medium-sized {\it HST} programme, the second effectively matches the depth of the available $B_{435}$ imaging in the GOODS fields, while the final deepest depth was chosen
to match (in terms of AB magnitudes) the depth of the existing $V_{606}$ and $i_{814}$ imaging in the UDS-CANDELS and COSMOS-CANDELS fields. 

\begin{figure*}
   \centering
   \includegraphics[width=18cm]{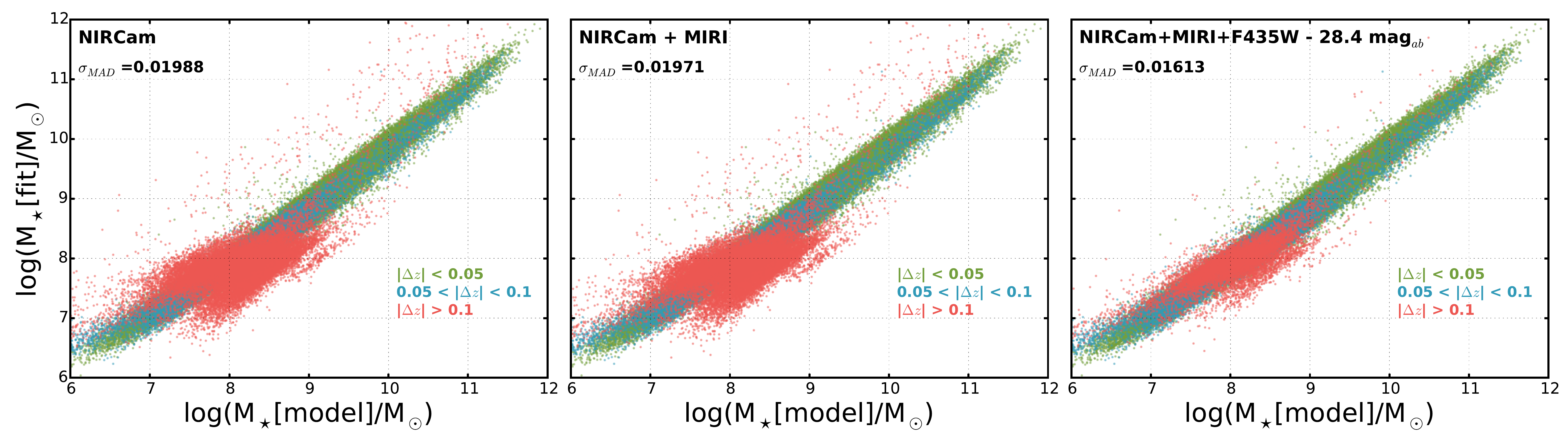}
   \caption{Reclaimed galaxy stellar mass versus input fiducial model mass for the galaxies in our {\it JWST}-detected EGG Sample, with NIRCam alone (Left), NIRCam with added MIRI F770W (Centre) and with the
     addition of deep {\it HST} ACS $B_{435}$ photometry reaching 28.4\,mag (5-$\sigma$) (Right). We colour code by photometric redshift offset ($|\Delta z|$) in three separate bins. The green points represent galaxies fitted with well-constrained photometric redshifts ($|\Delta z| < 0.05$), the blue points represent galaxies that have reasonably good fits to redshift ($0.05 < |\Delta z| < 0.1$) while the red points represent galaxies with poor photometric redshift estimations ($|\Delta z| > 0.1$).
     We also give the mean $|\Delta m|$ in each case, where $|\Delta m|$ is defined as 
$\log(M_*$[model]/${\rm M_{\odot}}$) $-$ $\log(M_*$[fit]/${\rm M_{\odot}})$.}\label{Fig:EGG_mm}
\end{figure*} 

The impact of this simulated progressive increase in $B_{435}$ imaging depth on the photometric redshifts delivered by {\sc Bagpipes} for the EGG {\it JWST} galaxy sample is shown in Fig.\,\ref{Fig:EGG_zz}.
There is a major impact on the number of incorrect redshifts at $z < 3.5$, and as a result the COR is predicted to decline from $\simeq 12$\% to $\simeq 4$\% simply by deepening the ACS $B_{435}$ imaging
by a factor of $\simeq 2$ in limiting flux density. The vertical arm feature representing $z \simeq 1$ galaxies with seriously over-estimated
photometric redshifts almost completely disappears as we now have a good enough detection of the flux blueward of 0.6$\mu$m. Previously, the poor detection allows for an incorrect confirmation of the Lyman-break pushing the posterior median of the photometric redshift distribution to higher redshifts.  We also see a substantial number of genuine $z \simeq 2 - 3.5$ galaxies have their previously under-estimated redshifts corrected by
better confirmation of the Lyman-break feature due to the deeper $B_{435}$ imaging. We see no real improvement beyond $z > 5$ as, for many of these galaxies, a more secure non-detection
in the $B$-band adds little to the information provided by the existing $V_{606}$ imaging given the location of the Lyman break.

The relative added value of either MIRI F770W or {\it HST} $B_{435}$ imaging is compared in Fig.\,\ref{Fig:EGG_zz_2}. Here we show the inferred photometric redshift versus the
fiducial input redshift for 3 separate filter combinations of photometric data. In producing the results shown in the left-hand panel we used {\it HST} $V_{606}$, {\it HST} $i_{814}$ and the
8 {\it JWST} NIRCam bands, in the middle panel we added MIRI F770W photometry, and for the right-hand panel we instead added deep (28.4\,mag.)
{\it HST} $B_{435}$ photometry. Consistent with the above results, we see a small improvement when we add the MIRI F770W parallel photometry,
with COR reduced by $\sim$2.3\%, while the addition of deep {\it HST} $B_{435}$ imaging decreases the COR by $\sim$10\%.

As already demonstrated for the example galaxy highlighted in Fig.\,\ref{fig:eg_n7} and Fig.\,\ref{fig:eg}, an error in photometric redshift is one of the key sources
of systematic error in reclaimed galaxy stellar mass. Consequently, catastrophic redshift errors of the type seen in Fig.\,\ref{Fig:EGG_zz} and Fig.\,\ref{Fig:EGG_zz_2} can have a serious
impact on the determination of galaxy mass functions. For example, a galaxy in the vertical arm feature focused at a true redshift of $z \simeq 1$ in Fig.\,\ref{Fig:EGG_zz} and Fig.\,\ref{Fig:EGG_zz_2}
could have its redshift over-estimated up to a value of $z_{phot} \simeq 5.3$. This alters the inferred galaxy stellar mass from $M_* = 10^{6.4}\,{\rm M_{\odot}}$ to $M_* = 10^{9.1}\,{\rm M_{\odot}}$.
Fig.\,\ref{Fig:EGG_mm} shows reclaimed stellar mass versus the fiducial model stellar mass for the galaxies in our EGG sample first with NIRCam only, then NIRCam with the addition of MIRI F770W imaging and finally with the
proposed deeper {\it HST} $B_{435}$ imaging data (reaching 28.4\,mag, 5-$\sigma$). We colour code each panel with 3 photometric redshift offsets bins: the green points represent galaxies with a good photometric redshift ($|\Delta z| < 0.05$), the blue points represent galaxies with reasonably well fitted photometric redshifts ($0.05 < |\Delta z| < 0.1$), while the red points represent galaxies with a poorly inferred photometric redshift ($|\Delta z| > 0.1$). The relationship is significantly tightened only with the addition of the deep $B$-band data where a large number of galaxies with true stellar masses in the range $M_* = 10^{7} - 10^{10}\,{\rm M_{\odot}}$
have their reclaimed masses restored to their true (i.e. input) values. This is primarily as a result of the reduction in catastrophic redshifts, where not only do we see a dramatic decrease in scatter as we add the $B_{435}$ imaging, but also the vast reduction in the number of galaxies with poor photometric redshift constraints ({\it i.e} reduction in the number of red points). The normalised change in stellar mass $|\Delta m|$ (where
$|\Delta m|$~=~$\log(M_*$[model]/${\rm M_{\odot}}$) $-$ $\log(M_*$[fit]/${\rm M_{\odot}}$) drops by 0.0336, reflecting the accuracy with which the stellar masses are reclaimed with
the addition of the deep $B_{435}$ imaging. Clearly an accurate determination of galaxy stellar masses is key to an effective confrontation between the results from
{\it JWST} galaxy surveys and the predictions of theoretical models and simulations. These results indicate that deep $B_{435}$ {\it HST} optical imaging in the CANDELS fields
is crucial for minimising systematic errors in the stellar masses of the galaxies that will be uncovered in deep {\it JWST} surveys.

\section{Summary and Conclusions}\label{sec6:sum_conclusion}

We have undertaken an analysis of simulated {\it JWST}+{\it HST} photometry
for a wide range of model galaxies, in order to explore the ability with which
the redshifts and physical properties of galaxies can be reclaimed from different
multi-band combinations of photometry. In particular we have explored the potential
value of extending the wavelength baseline of {\it JWST} NIRCam photometry to longer
wavelengths with {\it JWST} MIRI, and/or to shorter wavelengths with
additional {\it HST} ACS imaging. The former is of interest because, for moderate size
NIRCam surveys, MIRI data can be obtained in parallel, while the latter is of interest
because we still have the opportunity to utilise {\it HST} to
improve the available optical imaging in likely {\it JWST} extragalactic survey fields.

In this paper we ensure our results are fully realistic, because {\bf i)}
they are based on fully-developed observing strategies which are achievable with {\it      
JWST} in the selected survey fields; {\bf ii)} we assume
MIRI integration times that are achievable in parallel-mode observing, and {\bf iii)}
we use the latest information on available {\it HST} imaging in the survey
fields, while also exploring what improvements
can be achieved with feasible future {\it HST} large programs.

We created two different simulated galaxy samples, one (based on the models
of Bruzual \& Charlot 2003) designed to explore a wide range of possible galaxy properties
including very dust-obscured objects, and a second
(based on the the Empirical Galaxy Generator: EGG) that is designed to be representative
of the known galaxy population. The photometry for all these objects was created
and scattered given the latest knowledge of the preformance of the relevant instruments,
and the samples were then trimmed to only include significant detections (given our
adopted observing strategy). Finally we used the Bayesian SED-fitting code {\sc Bagpipes}
to attempt to reclaim the properties of the galaxies from the scattered photometry,
performing the analysis both with and without the inclusion of the MIRI and/or {\it        
HST} ACS data.

Our main findings are that {\bf i)} parallel MIRI 7.7\,$\mu$m imaging is of real
value (in some cases crucial), but only for a subset of galaxies at $z \simeq 4 - 7$
which have high
dust obscuration, $A_V > 2$; {\bf ii)} the current EGG simulations suggest that
such galaxies only constitute 0.5\% of the $z = 0 - 12$ galaxy population
that will be detected by {\it JWST} NIRCam surveys, but this could
easily be an under-estimate, given the known existence of galaxies in the
GOODS fields with $A_V \simeq 5$ (e.g. \citealp{Dunlop:2007}); {\bf iii)} while the MIRI photometry
also appears to help with the accurate recovery of the properties of some galaxies
at $z < 4$, it transpires that deep {\it HST} $B$-band photometry is of more
value (and is more efficient) for removing photometric redshift degeneracies for such
galaxies.

Given the importance of robustly quantifying the evolution of the high-mass
dusty galaxy population out to the highest redshifts, our results still favour
the use of MIRI parallels with deep NIRCam surveys, despite the fact that, for the vast
majority of the bluer galaxy population, the MIRI photometry struggles to add value to
the deeper NIRCam data. Our results also argue strongly for the acquisition
of additional deep public {\it HST} ACS imaging in the key CANDELS fields (e.g. UDS and COSMOS)
prior to the commencement of {\it JWST} operations, to ensure that the power of
deep extra-galactic imaging surveys with {\it JWST} is maximised.

Moreover, in closing we also note that, given the known existence of dusty galaxies such as the example highlighted in Figs. \ref{fig:eg_n7} and \ref{fig:eg}, even if such objects prove to be rare on the scale of the CANDELS fields, they will undoubtedly be revealed in wider area surveys ({\it e.g.} in on-going near-infrared or sub-mm/mm surveys). This means that deep pointed MIRI follow-up of such sources will be invaluable, both for resolving redshifts/mass ambiguities, and indeed potentially for delineating their mid-infrared SED's through multi-band MIRI imaging. 

\section*{Acknowledgments}
TK, JSD, RJM, ACC and FC acknowledge the support of STFC.
This work is based in part on observations made with the NASA/ESA Hubble Space Telescope,
which is operated by the Association of Universities for Research in Astronomy, Inc.,
under NASA contract NAS5-26555. This work is also based in part on observations made with the Spitzer Space Telescope,
which is operated by the Jet Propulsion Laboratory, California Institute of Technology under NASA contract 1407.

\appendix

\bibliographystyle{mnras}
\bibliography{refer}

\begin{thebibliography}{}
\makeatletter
\relax
\def\mn@urlcharsother{\let\do\@makeother \do\$\do\&\do\#\do\^\do\_\do\%\do\~}
\def\mn@doi{\begingroup\mn@urlcharsother \@ifnextchar [ {\mn@doi@}
  {\mn@doi@[]}}
\def\mn@doi@[#1]#2{\def\@tempa{#1}\ifx\@tempa\@empty \href
  {http://dx.doi.org/#2} {doi:#2}\else \href {http://dx.doi.org/#2} {#1}\fi
  \endgroup}
\def\mn@eprint#1#2{\mn@eprint@#1:#2::\@nil}
\def\mn@eprint@arXiv#1{\href {http://arxiv.org/abs/#1} {{\tt arXiv:#1}}}
\def\mn@eprint@dblp#1{\href {http://dblp.uni-trier.de/rec/bibtex/#1.xml}
  {dblp:#1}}
\def\mn@eprint@#1:#2:#3:#4\@nil{\def\@tempa {#1}\def\@tempb {#2}\def\@tempc
  {#3}\ifx \@tempc \@empty \let \@tempc \@tempb \let \@tempb \@tempa \fi \ifx
  \@tempb \@empty \def\@tempb {arXiv}\fi \@ifundefined
  {mn@eprint@\@tempb}{\@tempb:\@tempc}{\expandafter \expandafter \csname
  mn@eprint@\@tempb\endcsname \expandafter{\@tempc}}}

\bibitem[\protect\citeauthoryear{{Ashby} et~al.,}{{Ashby}
  et~al.}{2013}]{Ashby:2013}
{Ashby} M.~L.~N.,  et~al., 2013, \mn@doi [\apj] {10.1088/0004-637X/769/1/80},
  \href {http://cdsads.u-strasbg.fr/abs/2013ApJ...769...80A} {769, 80}

\bibitem[\protect\citeauthoryear{{Ashby} et~al.,}{{Ashby}
  et~al.}{2015}]{Ashby:2015}
{Ashby} M.~L.~N.,  et~al., 2015, \mn@doi [\apjs] {10.1088/0067-0049/218/2/33},
  \href {http://cdsads.u-strasbg.fr/abs/2015ApJS..218...33A} {218, 33}

\bibitem[\protect\citeauthoryear{{Ashby} et~al.,}{{Ashby}
  et~al.}{2018}]{Ashby:2018}
{Ashby} M.~L.~N.,  et~al., 2018, \mn@doi [\apjs] {10.3847/1538-4365/aad4fb},
  \href {http://cdsads.u-strasbg.fr/abs/2018ApJS..237...39A} {237, 39}

\bibitem[\protect\citeauthoryear{{Barrow}, {Wise}, {Norman}, {O'Shea}  \&
  {Xu}}{{Barrow} et~al.}{2017}]{Barrow:2017}
{Barrow} K.~S.~S.,  {Wise} J.~H.,  {Norman} M.~L.,  {O'Shea} B.~W.,   {Xu} H.,
  2017, \mn@doi [\mnras] {10.1093/mnras/stx1181}, \href
  {http://cdsads.u-strasbg.fr/abs/2017MNRAS.469.4863B} {469, 4863}

\bibitem[\protect\citeauthoryear{{Beckwith} et~al.,}{{Beckwith}
  et~al.}{2006}]{Beckwith:2006}
{Beckwith} S.~V.~W.,  et~al., 2006, \mn@doi [\aj] {10.1086/507302}, \href
  {http://adsabs.harvard.edu/abs/2006AJ....132.1729B} {132, 1729}

\bibitem[\protect\citeauthoryear{{Bisigello} et~al.,}{{Bisigello}
  et~al.}{2016}]{Bisigello:2016}
{Bisigello} L.,  et~al., 2016, \mn@doi [\apjs] {10.3847/0067-0049/227/2/19},
  \href {http://adsabs.harvard.edu/abs/2016ApJS..227...19B} {227, 19}

\bibitem[\protect\citeauthoryear{{Bisigello} et~al.,}{{Bisigello}
  et~al.}{2017}]{Bisigello:2017}
{Bisigello} L.,  et~al., 2017, \mn@doi [\apjs] {10.3847/1538-4365/aa7a14},
  \href {http://cdsads.u-strasbg.fr/abs/2017ApJS..231....3B} {231, 3}

\bibitem[\protect\citeauthoryear{{Bouchet} et~al.,}{{Bouchet}
  et~al.}{2015}]{Bouchet:2015}
{Bouchet} P.,  et~al., 2015, \mn@doi [\pasp] {10.1086/682254}, \href
  {http://adsabs.harvard.edu/abs/2015PASP..127..612B} {127, 612}

\bibitem[\protect\citeauthoryear{{Bourne} et~al.,}{{Bourne}
  et~al.}{2017}]{Bourne:2017}
{Bourne} N.,  et~al., 2017, \mn@doi [\mnras] {10.1093/mnras/stx031}, \href
  {http://cdsads.u-strasbg.fr/abs/2017MNRAS.467.1360B} {467, 1360}

\bibitem[\protect\citeauthoryear{{Bouwens} et~al.,}{{Bouwens}
  et~al.}{2011}]{Bouwens:2011}
{Bouwens} R.~J.,  et~al., 2011, \mn@doi [\apj] {10.1088/0004-637X/737/2/90},
  \href {http://cdsads.u-strasbg.fr/abs/2011ApJ...737...90B} {737, 90}

\bibitem[\protect\citeauthoryear{{Bouwens} et~al.,}{{Bouwens}
  et~al.}{2015}]{Bouwens:2015}
{Bouwens} R.~J.,  et~al., 2015, \mn@doi [\apj] {10.1088/0004-637X/803/1/34},
  \href {http://cdsads.u-strasbg.fr/abs/2015ApJ...803...34B} {803, 34}

\bibitem[\protect\citeauthoryear{{Bouwens}, {Oesch}, {Illingworth}, {Ellis}  \&
  {Stefanon}}{{Bouwens} et~al.}{2017}]{Bouwens:2017}
{Bouwens} R.~J.,  {Oesch} P.~A.,  {Illingworth} G.~D.,  {Ellis} R.~S.,
  {Stefanon} M.,  2017, \mn@doi [\apj] {10.3847/1538-4357/aa70a4}, \href
  {http://cdsads.u-strasbg.fr/abs/2017ApJ...843..129B} {843, 129}

\bibitem[\protect\citeauthoryear{{Bowler} et~al.,}{{Bowler}
  et~al.}{2014}]{Bowler:2014}
{Bowler} R.~A.~A.,  et~al., 2014, \mn@doi [\mnras] {10.1093/mnras/stu449},
  \href {http://cdsads.u-strasbg.fr/abs/2014MNRAS.440.2810B} {440, 2810}

\bibitem[\protect\citeauthoryear{{Bowler} et~al.,}{{Bowler}
  et~al.}{2015}]{Bowler:2015}
{Bowler} R.~A.~A.,  et~al., 2015, \mn@doi [\mnras] {10.1093/mnras/stv1403},
  \href {http://cdsads.u-strasbg.fr/abs/2015MNRAS.452.1817B} {452, 1817}

\bibitem[\protect\citeauthoryear{{Brammer} et~al.,}{{Brammer}
  et~al.}{2016}]{Brammer:2016}
{Brammer} G.~B.,  et~al., 2016, \mn@doi [\apjs] {10.3847/0067-0049/226/1/6},
  \href {http://cdsads.u-strasbg.fr/abs/2016ApJS..226....6B} {226, 6}

\bibitem[\protect\citeauthoryear{{Bruzual} \& {Charlot}}{{Bruzual} \&
  {Charlot}}{2003}]{B&C:2003}
{Bruzual} G.,  {Charlot} S.,  2003, \mn@doi [\mnras]
  {10.1046/j.1365-8711.2003.06897.x}, \href
  {http://adsabs.harvard.edu/abs/2003MNRAS.344.1000B} {344, 1000}

\bibitem[\protect\citeauthoryear{{Calzetti}, {Kinney}  \&
  {Storchi-Bergmann}}{{Calzetti} et~al.}{1994}]{Calzetti:1994}
{Calzetti} D.,  {Kinney} A.~L.,   {Storchi-Bergmann} T.,  1994, \mn@doi [\apj]
  {10.1086/174346}, \href {http://adsabs.harvard.edu/abs/1994ApJ...429..582C}
  {429, 582}

\bibitem[\protect\citeauthoryear{{Calzetti}, {Armus}, {Bohlin}, {Kinney},
  {Koornneef}  \& {Storchi-Bergmann}}{{Calzetti} et~al.}{2000}]{Calzetti:2000}
{Calzetti} D.,  {Armus} L.,  {Bohlin} R.~C.,  {Kinney} A.~L.,  {Koornneef} J.,
   {Storchi-Bergmann} T.,  2000, \mn@doi [\apj] {10.1086/308692}, \href
  {http://adsabs.harvard.edu/abs/2000ApJ...533..682C} {533, 682}

\bibitem[\protect\citeauthoryear{{Carnall}, {Leja}, {Johnson}, {McLure},
  {Dunlop}  \& {Conroy}}{{Carnall} et~al.}{2018a}]{Carnall_2:2018}
{Carnall} A.~C.,  {Leja} J.,  {Johnson} B.~D.,  {McLure} R.~J.,  {Dunlop}
  J.~S.,   {Conroy} C.,  2018a, arXiv e-prints, \href
  {http://cdsads.u-strasbg.fr/abs/2018arXiv181103635C} {}

\bibitem[\protect\citeauthoryear{{Carnall}, {McLure}, {Dunlop}  \&
  {Dav{\'e}}}{{Carnall} et~al.}{2018b}]{Carnall:2018}
{Carnall} A.~C.,  {McLure} R.~J.,  {Dunlop} J.~S.,   {Dav{\'e}} R.,  2018b,
  \mn@doi [\mnras] {10.1093/mnras/sty2169}, \href
  {http://cdsads.u-strasbg.fr/abs/2018MNRAS.480.4379C} {480, 4379}

\bibitem[\protect\citeauthoryear{{Cirasuolo}, {McLure}, {Dunlop}, {Almaini},
  {Foucaud}  \& {Simpson}}{{Cirasuolo} et~al.}{2010}]{Cirasuolo:2010}
{Cirasuolo} M.,  {McLure} R.~J.,  {Dunlop} J.~S.,  {Almaini} O.,  {Foucaud} S.,
    {Simpson} C.,  2010, \mn@doi [\mnras] {10.1111/j.1365-2966.2009.15710.x},
  \href {http://cdsads.u-strasbg.fr/abs/2010MNRAS.401.1166C} {401, 1166}

\bibitem[\protect\citeauthoryear{{Cowley}, {Baugh}, {Cole}, {Frenk}  \&
  {Lacey}}{{Cowley} et~al.}{2018}]{Cowley:2018}
{Cowley} W.~I.,  {Baugh} C.~M.,  {Cole} S.,  {Frenk} C.~S.,   {Lacey} C.~G.,
  2018, \mn@doi [\mnras] {10.1093/mnras/stx2897}, \href
  {http://adsabs.harvard.edu/abs/2018MNRAS.474.2352C} {474, 2352}

\bibitem[\protect\citeauthoryear{{Cullen}, {McLure}, {Khochfar}, {Dunlop}  \&
  {Dalla Vecchia}}{{Cullen} et~al.}{2017}]{Cullen:2017}
{Cullen} F.,  {McLure} R.~J.,  {Khochfar} S.,  {Dunlop} J.~S.,   {Dalla
  Vecchia} C.,  2017, \mn@doi [\mnras] {10.1093/mnras/stx1451}, \href
  {http://adsabs.harvard.edu/abs/2017MNRAS.470.3006C} {470, 3006}

\bibitem[\protect\citeauthoryear{{Dahlen} et~al.,}{{Dahlen}
  et~al.}{2013}]{Dahlen:2013}
{Dahlen} T.,  et~al., 2013, \mn@doi [\apj] {10.1088/0004-637X/775/2/93}, \href
  {http://cdsads.u-strasbg.fr/abs/2013ApJ...775...93D} {775, 93}

\bibitem[\protect\citeauthoryear{{Damen} et~al.,}{{Damen}
  et~al.}{2011}]{Damen:2011}
{Damen} M.,  et~al., 2011, \mn@doi [\apj] {10.1088/0004-637X/727/1/1}, \href
  {http://cdsads.u-strasbg.fr/abs/2011ApJ...727....1D} {727, 1}

\bibitem[\protect\citeauthoryear{{Davidzon} et~al.,}{{Davidzon}
  et~al.}{2017}]{Davidzon:2017}
{Davidzon} I.,  et~al., 2017, \mn@doi [\aap] {10.1051/0004-6361/201730419},
  \href {http://cdsads.u-strasbg.fr/abs/2017A%26A...605A..70D} {605, A70}

\bibitem[\protect\citeauthoryear{{Doyon} et~al.,}{{Doyon}
  et~al.}{2012}]{Doyon:2012}
{Doyon} R.,  et~al., 2012, in Space Telescopes and Instrumentation 2012:
  Optical, Infrared, and Millimeter Wave. p. 84422R, \mn@doi{10.1117/12.926578}

\bibitem[\protect\citeauthoryear{{Dunlop}}{{Dunlop}}{2013}]{Dunlop_Review:2013}
{Dunlop} J.~S.,  2013, in {Wiklind} T.,  {Mobasher} B.,   {Bromm} V.,  eds,
  Astrophysics and Space Science Library Vol. 396, The First Galaxies. p.~223
  (\mn@eprint {arXiv} {1205.1543}), \mn@doi{10.1007/978-3-642-32362-1_5}

\bibitem[\protect\citeauthoryear{{Dunlop}, {Cirasuolo}  \& {McLure}}{{Dunlop}
  et~al.}{2007}]{Dunlop:2007}
{Dunlop} J.~S.,  {Cirasuolo} M.,   {McLure} R.~J.,  2007, \mn@doi [\mnras]
  {10.1111/j.1365-2966.2007.11453.x}, \href
  {http://cdsads.u-strasbg.fr/abs/2007MNRAS.376.1054D} {376, 1054}

\bibitem[\protect\citeauthoryear{{Dunlop} et~al.,}{{Dunlop}
  et~al.}{2017}]{Dunlop:2017}
{Dunlop} J.~S.,  et~al., 2017, \mn@doi [\mnras] {10.1093/mnras/stw3088}, \href
  {http://cdsads.u-strasbg.fr/abs/2017MNRAS.466..861D} {466, 861}

\bibitem[\protect\citeauthoryear{{Ellis} et~al.,}{{Ellis}
  et~al.}{2013}]{Ellis:2013}
{Ellis} R.~S.,  et~al., 2013, \mn@doi [\apjl] {10.1088/2041-8205/763/1/L7},
  \href {http://adsabs.harvard.edu/abs/2013ApJ...763L...7E} {763, L7}

\bibitem[\protect\citeauthoryear{{Feng}, {Di-Matteo}, {Croft}, {Bird},
  {Battaglia}  \& {Wilkins}}{{Feng} et~al.}{2016}]{Feng:2016}
{Feng} Y.,  {Di-Matteo} T.,  {Croft} R.~A.,  {Bird} S.,  {Battaglia} N.,
  {Wilkins} S.,  2016, \mn@doi [\mnras] {10.1093/mnras/stv2484}, \href
  {http://adsabs.harvard.edu/abs/2016MNRAS.455.2778F} {455, 2778}

\bibitem[\protect\citeauthoryear{{Ferland} et~al.,}{{Ferland}
  et~al.}{2017}]{Ferland:2017}
{Ferland} G.~J.,  et~al., 2017, \rmxaa, \href
  {http://adsabs.harvard.edu/abs/2017RMxAA..53..385F} {53, 385}

\bibitem[\protect\citeauthoryear{{Finkelstein}}{{Finkelstein}}{2016}]{Finkelstein:2016}
{Finkelstein} S.~L.,  2016, \mn@doi [\pasa] {10.1017/pasa.2016.26}, \href
  {http://cdsads.u-strasbg.fr/abs/2016PASA...33...37F} {33, e037}

\bibitem[\protect\citeauthoryear{{Finkelstein}, {Dunlop}, {Le Fevre}  \&
  {Wilkins}}{{Finkelstein} et~al.}{2015}]{Finkelstein:2015}
{Finkelstein} S.~L.,  {Dunlop} J.,  {Le Fevre} O.,   {Wilkins} S.,  2015,
  preprint, \href {http://cdsads.u-strasbg.fr/abs/2015arXiv151204530F} {}
  (\mn@eprint {arXiv} {1512.04530})

\bibitem[\protect\citeauthoryear{{Finkelstein} et~al.,}{{Finkelstein}
  et~al.}{2017}]{Finkelstein:2017}
{Finkelstein} S.,  et~al., 2017, {The Cosmic Evolution Early Release Science
  (CEERS) Survey}, JWST Proposal ID 1345. Cycle 0 Early Release Science

\bibitem[\protect\citeauthoryear{{Finlator}, {Oppenheimer}  \&
  {Dav{\'e}}}{{Finlator} et~al.}{2011}]{Finlator:2011}
{Finlator} K.,  {Oppenheimer} B.~D.,   {Dav{\'e}} R.,  2011, \mn@doi [\mnras]
  {10.1111/j.1365-2966.2010.17554.x}, \href
  {http://adsabs.harvard.edu/abs/2011MNRAS.410.1703F} {410, 1703}

\bibitem[\protect\citeauthoryear{{Fontana} et~al.,}{{Fontana}
  et~al.}{2006}]{Fontana:2006}
{Fontana} A.,  et~al., 2006, \mn@doi [\aap] {10.1051/0004-6361:20065475}, \href
  {http://cdsads.u-strasbg.fr/abs/2006A%26A...459..745F} {459, 745}

\bibitem[\protect\citeauthoryear{{Fontana} et~al.,}{{Fontana}
  et~al.}{2009}]{Fontana:2009}
{Fontana} A.,  et~al., 2009, \mn@doi [\aap] {10.1051/0004-6361/200911650},
  \href {http://cdsads.u-strasbg.fr/abs/2009A%26A...501...15F} {501, 15}

\bibitem[\protect\citeauthoryear{{Fontana} et~al.,}{{Fontana}
  et~al.}{2014}]{Fontana:2014}
{Fontana} A.,  et~al., 2014, \mn@doi [\aap] {10.1051/0004-6361/201423543},
  \href {http://cdsads.u-strasbg.fr/abs/2014A%26A...570A..11F} {570, A11}

\bibitem[\protect\citeauthoryear{{Furusawa} et~al.,}{{Furusawa}
  et~al.}{2008}]{Furusawa:2008}
{Furusawa} H.,  et~al., 2008, \mn@doi [\apjs] {10.1086/527321}, \href
  {http://cdsads.u-strasbg.fr/abs/2008ApJS..176....1F} {176, 1}

\bibitem[\protect\citeauthoryear{{Furusawa} et~al.,}{{Furusawa}
  et~al.}{2016}]{Furusawa:2016}
{Furusawa} H.,  et~al., 2016, \mn@doi [\apj] {10.3847/0004-637X/822/1/46},
  \href {http://cdsads.u-strasbg.fr/abs/2016ApJ...822...46F} {822, 46}

\bibitem[\protect\citeauthoryear{Gardner et~al.,}{Gardner
  et~al.}{2006}]{Gardner:2006}
Gardner J.~P.,  et~al., 2006, \mn@doi [Space Science Reviews]
  {10.1007/s11214-006-8315-7}, 123, 485

\bibitem[\protect\citeauthoryear{{Giavalisco} et~al.,}{{Giavalisco}
  et~al.}{2004}]{Giavalisco:2004}
{Giavalisco} M.,  et~al., 2004, \mn@doi [\apjl] {10.1086/379232}, \href
  {http://cdsads.u-strasbg.fr/abs/2004ApJ...600L..93G} {600, L93}

\bibitem[\protect\citeauthoryear{{Glasse} et~al.,}{{Glasse}
  et~al.}{2015}]{Glasse:2015}
{Glasse} A.,  et~al., 2015, \mn@doi [\pasp] {10.1086/682259}, \href
  {http://adsabs.harvard.edu/abs/2015PASP..127..686G} {127, 686}

\bibitem[\protect\citeauthoryear{{Glazebrook} et~al.,}{{Glazebrook}
  et~al.}{2017}]{Glazebrook:2017}
{Glazebrook} K.,  et~al., 2017, \mn@doi [\nat] {10.1038/nature21680}, \href
  {http://cdsads.u-strasbg.fr/abs/2017Natur.544...71G} {544, 71}

\bibitem[\protect\citeauthoryear{{Grogin} et~al.,}{{Grogin}
  et~al.}{2011}]{Grogin:2011}
{Grogin} N.~A.,  et~al., 2011, \mn@doi [\apjs] {10.1088/0067-0049/197/2/35},
  \href {http://adsabs.harvard.edu/abs/2011ApJS..197...35G} {197, 35}

\bibitem[\protect\citeauthoryear{{Hartley}, {Almaini}  \& {Foucaud}}{{Hartley}
  et~al.}{2013}]{Hartley:2013}
{Hartley} W.~G.,  {Almaini} O.,   {Foucaud} S.,  2013, in {Adamson} A.,
  {Davies} J.,   {Robson} I.,  eds,  Astrophysics and Space Science Proceedings
  Vol. 37, Thirty Years of Astronomical Discovery with UKIRT. p.~309,
  \mn@doi{10.1007/978-94-007-7432-2_29}

\bibitem[\protect\citeauthoryear{{Horner} \& {Rieke}}{{Horner} \&
  {Rieke}}{2004}]{Horner:2004}
{Horner} S.~D.,  {Rieke} M.~J.,  2004, in {Mather} J.~C.,  ed.,  \procspie Vol.
  5487, Optical, Infrared, and Millimeter Space Telescopes. pp 628--634,
  \mn@doi{10.1117/12.552281}

\bibitem[\protect\citeauthoryear{{Ilbert} et~al.,}{{Ilbert}
  et~al.}{2013}]{Ilbert:2013}
{Ilbert} O.,  et~al., 2013, \mn@doi [\aap] {10.1051/0004-6361/201321100}, \href
  {http://cdsads.u-strasbg.fr/abs/2013A%26A...556A..55I} {556, A55}

\bibitem[\protect\citeauthoryear{{Illingworth} et~al.,}{{Illingworth}
  et~al.}{2013}]{Illingworth:2013}
{Illingworth} G.~D.,  et~al., 2013, \mn@doi [\apjs]
  {10.1088/0067-0049/209/1/6}, \href
  {http://adsabs.harvard.edu/abs/2013ApJS..209....6I} {209, 6}

\bibitem[\protect\citeauthoryear{{Inoue}, {Shimizu}, {Iwata}  \&
  {Tanaka}}{{Inoue} et~al.}{2014}]{Inoue:2014}
{Inoue} A.~K.,  {Shimizu} I.,  {Iwata} I.,   {Tanaka} M.,  2014, \mn@doi
  [\mnras] {10.1093/mnras/stu936}, \href
  {http://adsabs.harvard.edu/abs/2014MNRAS.442.1805I} {442, 1805}

\bibitem[\protect\citeauthoryear{{Ishigaki}, {Kawamata}, {Ouchi}, {Oguri},
  {Shimasaku}  \& {Ono}}{{Ishigaki} et~al.}{2018}]{Ishigaki:2018}
{Ishigaki} M.,  {Kawamata} R.,  {Ouchi} M.,  {Oguri} M.,  {Shimasaku} K.,
  {Ono} Y.,  2018, \mn@doi [\apj] {10.3847/1538-4357/aaa544}, \href
  {http://cdsads.u-strasbg.fr/abs/2018ApJ...854...73I} {854, 73}

\bibitem[\protect\citeauthoryear{{Jarvis} et~al.,}{{Jarvis}
  et~al.}{2013}]{Jarvis:2013}
{Jarvis} M.~J.,  et~al., 2013, \mn@doi [\mnras] {10.1093/mnras/sts118}, \href
  {http://cdsads.u-strasbg.fr/abs/2013MNRAS.428.1281J} {428, 1281}

\bibitem[\protect\citeauthoryear{{Kendrew} et~al.,}{{Kendrew}
  et~al.}{2015}]{Kendrew:2015}
{Kendrew} S.,  et~al., 2015, \mn@doi [\pasp] {10.1086/682255}, \href
  {http://adsabs.harvard.edu/abs/2015PASP..127..623K} {127, 623}

\bibitem[\protect\citeauthoryear{{Koprowski} et~al.,}{{Koprowski}
  et~al.}{2018}]{Koprowski:2018}
{Koprowski} M.~P.,  et~al., 2018, \mn@doi [\mnras] {10.1093/mnras/sty1527},
  \href {http://cdsads.u-strasbg.fr/abs/2018MNRAS.479.4355K} {479, 4355}

\bibitem[\protect\citeauthoryear{{Labb{\'e}} et~al.,}{{Labb{\'e}}
  et~al.}{2013}]{Labbe:2013}
{Labb{\'e}} I.,  et~al., 2013, \mn@doi [\apjl] {10.1088/2041-8205/777/2/L19},
  \href {http://cdsads.u-strasbg.fr/abs/2013ApJ...777L..19L} {777, L19}

\bibitem[\protect\citeauthoryear{{Labb{\'e}} et~al.,}{{Labb{\'e}}
  et~al.}{2015}]{Labbe:2015}
{Labb{\'e}} I.,  et~al., 2015, \mn@doi [\apjs] {10.1088/0067-0049/221/2/23},
  \href {http://cdsads.u-strasbg.fr/abs/2015ApJS..221...23L} {221, 23}

\bibitem[\protect\citeauthoryear{{Lacey} et~al.,}{{Lacey}
  et~al.}{2016}]{Lacey:2016}
{Lacey} C.~G.,  et~al., 2016, \mn@doi [\mnras] {10.1093/mnras/stw1888}, \href
  {http://adsabs.harvard.edu/abs/2016MNRAS.462.3854L} {462, 3854}

\bibitem[\protect\citeauthoryear{{Laigle} et~al.,}{{Laigle}
  et~al.}{2016}]{Laigle:2016}
{Laigle} C.,  et~al., 2016, \mn@doi [\apjs] {10.3847/0067-0049/224/2/24}, \href
  {http://cdsads.u-strasbg.fr/abs/2016ApJS..224...24L} {224, 24}

\bibitem[\protect\citeauthoryear{{Lawrence} et~al.,}{{Lawrence}
  et~al.}{2007}]{Lawrence:2007}
{Lawrence} A.,  et~al., 2007, \mn@doi [\mnras]
  {10.1111/j.1365-2966.2007.12040.x}, \href
  {http://cdsads.u-strasbg.fr/abs/2007MNRAS.379.1599L} {379, 1599}

\bibitem[\protect\citeauthoryear{{Lotz} et~al.,}{{Lotz}
  et~al.}{2017}]{Lotz:2017}
{Lotz} J.~M.,  et~al., 2017, \mn@doi [\apj] {10.3847/1538-4357/837/1/97}, \href
  {http://cdsads.u-strasbg.fr/abs/2017ApJ...837...97L} {837, 97}

\bibitem[\protect\citeauthoryear{{Lovell} et~al.,}{{Lovell}
  et~al.}{2018}]{Lovell:2018}
{Lovell} M.~R.,  et~al., 2018, arXiv e-prints, \href
  {http://cdsads.u-strasbg.fr/abs/2018arXiv181005168L} {}

\bibitem[\protect\citeauthoryear{{Madau}}{{Madau}}{1995}]{Madau:1995}
{Madau} P.,  1995, \mn@doi [\apj] {10.1086/175332}, \href
  {http://adsabs.harvard.edu/abs/1995ApJ...441...18M} {441, 18}

\bibitem[\protect\citeauthoryear{{McCracken} et~al.,}{{McCracken}
  et~al.}{2012}]{McCracken:2012}
{McCracken} H.~J.,  et~al., 2012, \mn@doi [\aap] {10.1051/0004-6361/201219507},
  \href {http://cdsads.u-strasbg.fr/abs/2012A%26A...544A.156M} {544, A156}

\bibitem[\protect\citeauthoryear{{McLeod}, {McLure}, {Dunlop}, {Robertson},
  {Ellis}  \& {Targett}}{{McLeod} et~al.}{2015}]{McLeod:2015}
{McLeod} D.~J.,  {McLure} R.~J.,  {Dunlop} J.~S.,  {Robertson} B.~E.,  {Ellis}
  R.~S.,   {Targett} T.~A.,  2015, \mn@doi [\mnras] {10.1093/mnras/stv780},
  \href {http://cdsads.u-strasbg.fr/abs/2015MNRAS.450.3032M} {450, 3032}

\bibitem[\protect\citeauthoryear{{McLeod}, {McLure}  \& {Dunlop}}{{McLeod}
  et~al.}{2016}]{McLeod:2016}
{McLeod} D.~J.,  {McLure} R.~J.,   {Dunlop} J.~S.,  2016, \mn@doi [\mnras]
  {10.1093/mnras/stw904}, \href
  {http://cdsads.u-strasbg.fr/abs/2016MNRAS.459.3812M} {459, 3812}

\bibitem[\protect\citeauthoryear{{McLure}, {Dunlop}, {Cirasuolo}, {Koekemoer},
  {Sabbi}, {Stark}, {Targett}  \& {Ellis}}{{McLure} et~al.}{2010}]{McLure:2010}
{McLure} R.~J.,  {Dunlop} J.~S.,  {Cirasuolo} M.,  {Koekemoer} A.~M.,  {Sabbi}
  E.,  {Stark} D.~P.,  {Targett} T.~A.,   {Ellis} R.~S.,  2010, \mn@doi
  [\mnras] {10.1111/j.1365-2966.2009.16176.x}, \href
  {http://adsabs.harvard.edu/abs/2010MNRAS.403..960M} {403, 960}

\bibitem[\protect\citeauthoryear{{McLure} et~al.,}{{McLure}
  et~al.}{2013}]{McLure:2013}
{McLure} R.~J.,  et~al., 2013, \mn@doi [\mnras] {10.1093/mnras/stt627}, \href
  {http://adsabs.harvard.edu/abs/2013MNRAS.432.2696M} {432, 2696}

\bibitem[\protect\citeauthoryear{{McLure} et~al.,}{{McLure}
  et~al.}{2018a}]{McLure_2:2018}
{McLure} R.~J.,  et~al., 2018a, \mn@doi [\mnras] {10.1093/mnras/sty522}, \href
  {http://cdsads.u-strasbg.fr/abs/2018MNRAS.476.3991M} {476, 3991}

\bibitem[\protect\citeauthoryear{{McLure} et~al.,}{{McLure}
  et~al.}{2018b}]{McLure:2018}
{McLure} R.~J.,  et~al., 2018b, \mn@doi [\mnras] {10.1093/mnras/sty1213}, \href
  {http://adsabs.harvard.edu/abs/2018MNRAS.479...25M} {479, 25}

\bibitem[\protect\citeauthoryear{{Merlin} et~al.,}{{Merlin}
  et~al.}{2018}]{Merlin:2018}
{Merlin} E.,  et~al., 2018, \mn@doi [\mnras] {10.1093/mnras/stx2385}, \href
  {http://cdsads.u-strasbg.fr/abs/2018MNRAS.473.2098M} {473, 2098}

\bibitem[\protect\citeauthoryear{{Muzzin} et~al.,}{{Muzzin}
  et~al.}{2013}]{Muzzin:2013}
{Muzzin} A.,  et~al., 2013, \mn@doi [\apj] {10.1088/0004-637X/777/1/18}, \href
  {http://cdsads.u-strasbg.fr/abs/2013ApJ...777...18M} {777, 18}

\bibitem[\protect\citeauthoryear{{Nakajima}, {Ellis}, {Iwata}, {Inoue},
  {Kusakabe}, {Ouchi}  \& {Robertson}}{{Nakajima} et~al.}{2016}]{Nakajima:2016}
{Nakajima} K.,  {Ellis} R.~S.,  {Iwata} I.,  {Inoue} A.~K.,  {Kusakabe} H.,
  {Ouchi} M.,   {Robertson} B.~E.,  2016, \mn@doi [\apjl]
  {10.3847/2041-8205/831/1/L9}, \href
  {http://cdsads.u-strasbg.fr/abs/2016ApJ...831L...9N} {831, L9}

\bibitem[\protect\citeauthoryear{{Oesch}, {Bouwens}, {Illingworth}, {Labb{\'e}}
   \& {Stefanon}}{{Oesch} et~al.}{2018}]{Oesch:2018}
{Oesch} P.~A.,  {Bouwens} R.~J.,  {Illingworth} G.~D.,  {Labb{\'e}} I.,
  {Stefanon} M.,  2018, \mn@doi [\apj] {10.3847/1538-4357/aab03f}, \href
  {http://cdsads.u-strasbg.fr/abs/2018ApJ...855..105O} {855, 105}

\bibitem[\protect\citeauthoryear{{Oke}}{{Oke}}{1974}]{Oke:1974}
{Oke} J.~B.,  1974, \mn@doi [\apjs] {10.1086/190287}, \href
  {http://cdsads.u-strasbg.fr/abs/1974ApJS...27...21O} {27, 21}

\bibitem[\protect\citeauthoryear{{Oke} \& {Gunn}}{{Oke} \&
  {Gunn}}{1983}]{Oke:1983}
{Oke} J.~B.,  {Gunn} J.~E.,  1983, \mn@doi [\apj] {10.1086/160817}, \href
  {http://cdsads.u-strasbg.fr/abs/1983ApJ...266..713O} {266, 713}

\bibitem[\protect\citeauthoryear{{Paardekooper}, {Khochfar}  \& {Dalla
  Vecchia}}{{Paardekooper} et~al.}{2013}]{Paardekooper:2013}
{Paardekooper} J.-P.,  {Khochfar} S.,   {Dalla Vecchia} C.,  2013, \mn@doi
  [\mnras] {10.1093/mnrasl/sls032}, \href
  {http://adsabs.harvard.edu/abs/2013MNRAS.429L..94P} {429, L94}

\bibitem[\protect\citeauthoryear{{Parsa}, {Dunlop}, {McLure}  \&
  {Mortlock}}{{Parsa} et~al.}{2016}]{Parsa:2016}
{Parsa} S.,  {Dunlop} J.~S.,  {McLure} R.~J.,   {Mortlock} A.,  2016, \mn@doi
  [\mnras] {10.1093/mnras/stv2857}, \href
  {http://cdsads.u-strasbg.fr/abs/2016MNRAS.456.3194P} {456, 3194}

\bibitem[\protect\citeauthoryear{{Pontoppidan} et~al.,}{{Pontoppidan}
  et~al.}{2016}]{Pandeia:2017}
{Pontoppidan} K.~M.,  et~al., 2016, \mn@doi [\mnras] {10.1117/12.2231768},
  \href {http://adsabs.harvard.edu/abs/2016SPIE.9910E..16P} {9910, 991016}

\bibitem[\protect\citeauthoryear{{Posselt}, {Holota}, {Kulinyak}, {Kling},
  {Kutscheid}, {Le Fevre}, {Prieto}  \& {Ferruit}}{{Posselt}
  et~al.}{2004}]{Posselt:2004}
{Posselt} W.,  {Holota} W.,  {Kulinyak} E.,  {Kling} G.,  {Kutscheid} T.,  {Le
  Fevre} O.,  {Prieto} E.,   {Ferruit} P.,  2004, in {Mather} J.~C.,  ed.,
  \procspie Vol. 5487, Optical, Infrared, and Millimeter Space Telescopes. pp
  688--697, \mn@doi{10.1117/12.555659}

\bibitem[\protect\citeauthoryear{{Rieke} et~al.,}{{Rieke}
  et~al.}{2015}]{Rieke_MIRI:2015}
{Rieke} G.~H.,  et~al., 2015, \mn@doi [\pasp] {10.1086/682252}, \href
  {http://adsabs.harvard.edu/abs/2015PASP..127..584R} {127, 584}

\bibitem[\protect\citeauthoryear{{Sanders} et~al.,}{{Sanders}
  et~al.}{2007}]{Sanders:2007}
{Sanders} D.~B.,  et~al., 2007, \mn@doi [\apjs] {10.1086/517885}, \href
  {http://cdsads.u-strasbg.fr/abs/2007ApJS..172...86S} {172, 86}

\bibitem[\protect\citeauthoryear{{Santini} et~al.,}{{Santini}
  et~al.}{2015}]{Santini:2015}
{Santini} P.,  et~al., 2015, \mn@doi [\apj] {10.1088/0004-637X/801/2/97}, \href
  {http://cdsads.u-strasbg.fr/abs/2015ApJ...801...97S} {801, 97}

\bibitem[\protect\citeauthoryear{{Schaye} et~al.,}{{Schaye}
  et~al.}{2010}]{Schaye:2010}
{Schaye} J.,  et~al., 2010, \mn@doi [\mnras]
  {10.1111/j.1365-2966.2009.16029.x}, \href
  {http://adsabs.harvard.edu/abs/2010MNRAS.402.1536S} {402, 1536}

\bibitem[\protect\citeauthoryear{{Schreiber} et~al.,}{{Schreiber}
  et~al.}{2017}]{Corentin:2017}
{Schreiber} C.,  et~al., 2017, \mn@doi [\aap] {10.1051/0004-6361/201629123},
  \href {http://adsabs.harvard.edu/abs/2017A%26A...602A..96S} {602, A96}

\bibitem[\protect\citeauthoryear{{Scoville} et~al.,}{{Scoville}
  et~al.}{2007}]{Scoville:2007}
{Scoville} N.,  et~al., 2007, \mn@doi [\apjs] {10.1086/516580}, \href
  {http://adsabs.harvard.edu/abs/2007ApJS..172...38S} {172, 38}

\bibitem[\protect\citeauthoryear{{Simpson} et~al.,}{{Simpson}
  et~al.}{2014}]{Simpson:2014}
{Simpson} J.~M.,  et~al., 2014, \mn@doi [\apj] {10.1088/0004-637X/788/2/125},
  \href {http://cdsads.u-strasbg.fr/abs/2014ApJ...788..125S} {788, 125}

\bibitem[\protect\citeauthoryear{{Smit} et~al.,}{{Smit}
  et~al.}{2015}]{Smit:2015}
{Smit} R.,  et~al., 2015, \mn@doi [\apj] {10.1088/0004-637X/801/2/122}, \href
  {http://cdsads.u-strasbg.fr/abs/2015ApJ...801..122S} {801, 122}

\bibitem[\protect\citeauthoryear{{Stark}}{{Stark}}{2016}]{Stark:2016}
{Stark} D.~P.,  2016, \mn@doi [\araa] {10.1146/annurev-astro-081915-023417},
  \href {http://cdsads.u-strasbg.fr/abs/2016ARA%26A..54..761S} {54, 761}

\bibitem[\protect\citeauthoryear{{Taniguchi} et~al.,}{{Taniguchi}
  et~al.}{2007}]{Taniguchi:2007}
{Taniguchi} Y.,  et~al., 2007, \mn@doi [\apjs] {10.1086/516596}, \href
  {http://cdsads.u-strasbg.fr/abs/2007ApJS..172....9T} {172, 9}

\bibitem[\protect\citeauthoryear{{Tomczak} et~al.,}{{Tomczak}
  et~al.}{2014}]{Tomczak:2014}
{Tomczak} A.~R.,  et~al., 2014, \mn@doi [\apj] {10.1088/0004-637X/783/2/85},
  \href {http://cdsads.u-strasbg.fr/abs/2014ApJ...783...85T} {783, 85}

\bibitem[\protect\citeauthoryear{{Wilkins}, {Feng}, {Di Matteo}, {Croft},
  {Lovell}  \& {Waters}}{{Wilkins} et~al.}{2017}]{Wilkins:2017}
{Wilkins} S.~M.,  {Feng} Y.,  {Di Matteo} T.,  {Croft} R.,  {Lovell} C.~C.,
  {Waters} D.,  2017, \mn@doi [\mnras] {10.1093/mnras/stx841}, \href
  {http://adsabs.harvard.edu/abs/2017MNRAS.469.2517W} {469, 2517}

\bibitem[\protect\citeauthoryear{{Williams} et~al.,}{{Williams}
  et~al.}{2018}]{Williams:2018}
{Williams} C.~C.,  et~al., 2018, \mn@doi [\apjs] {10.3847/1538-4365/aabcbb},
  \href {http://adsabs.harvard.edu/abs/2018ApJS..236...33W} {236, 33}

\bibitem[\protect\citeauthoryear{{Yung}, {Somerville}, {Finkelstein}, {Popping}
   \& {Dav{\'e}}}{{Yung} et~al.}{2019}]{Yung:2019}
{Yung} L.~Y.~A.,  {Somerville} R.~S.,  {Finkelstein} S.~L.,  {Popping} G.,
  {Dav{\'e}} R.,  2019, \mn@doi [\mnras] {10.1093/mnras/sty3241}, \href
  {http://cdsads.u-strasbg.fr/abs/2019MNRAS.483.2983Y} {483, 2983}

\makeatother
\end{thebibliography}

\section{MIRI simulator}

The 25-m$^2$ collecting mirror of {\it JWST} provides the potential for huge gains in sensitivity for a mid-infrared telescope. This, coupled with the cold radiative environment ($\simeq 40$\,K) 
of {\it JWST} will allow MIRI to excel in a relatively unexplored region of the electromagnetic spectrum for extra-galactic science.  
As MIRI covers a large spectral range, care must be taken to track the change in dominating background factors as a function of wavelength. 
At short wavelengths, the Zodiacal dust concentrated in the ecliptic plane of our solar system dominates the background noise. However, as one moves to 
longer wavelengths the general thermal background increases by a few orders of magnitude.  The thermal background can be simply modelled as the sum of the six thermal emission sources expected when observing with MIRI, equation \ref{Equation:BB}, where each background source has a different effective temperature and emissivity ($\epsilon$, quantifying the efficiency of emission/absorption relative to a perfect black-body):

\begin{equation}
B_{\nu}= \epsilon\dfrac{2h\nu^3}{c^2} \dfrac{1}{e^{{h\nu}/{kT}}-1}
\label{Equation:BB}
\end{equation}

Table\, \ref{table:Greybody} presents a breakdown of the different contributions to the thermal background emission that MIRI will be exposed to when operating. Components A and B take into account the scattered and emissive components of the Zodiacal dust spectrum, while C to F account for stray light in the telescope. Component G is representative of a `high background', simulating double that currently expected in the optical system. When these components are plotted, the problems encountered when venturing into the longer wavelengths accessible with MIRI
are clear: the thermal background increases by several orders of magnitude, reducing the observational abilities of the instrument. At $\lambda > 14\,\mu$m, the background noise is dominated by the {\it JWST}  components and the thermal emission they produce when the telescope is in operation. 

To quantify the background noise an analytical integration can be performed over the grey-body profiles in the across the specified wavelength range, 
for example, the range of photometric bands for MIRI imaging, or the separate MRS bands and channels when undertaking spectroscopic observations. 
It is important to integrate across these channels separately, as a number of variables depend on the wavelength. The quantum efficiency, $\tau_{\lambda}$, and optical transmission $\eta_{\lambda}$ vary on small scales, and therefore we can use the mean values rather than integrating a designated profile. The Photon Conversion Efficiency (PCE: the number of detected electrons per incident photon) is the product of $\tau_{\lambda}$ and $\eta_{\lambda}$. The PCE stays relatively constant over small wavelength ranges, varying by about 3\% within the separate passbands. These grey-body profiles should be integrated individually to ensure the background in each channel is accurately calculated.

\begin{equation}
i_{bgd} = \Omega_{pix}\tau_{tel}\tau_{EOL}\int_{\lambda_1}^{\lambda_2}P_{bgd}(\lambda)\tau_{\lambda}\eta{_\lambda}d\lambda
\label{Equation:BgdPhotocurrent}
\end{equation}

$\Omega_{pix}$,  $\tau_{tel}$ and $\tau_{EOL}$ represent the solid angle field of view from each pixel and two forms of transmission efficiency respectively. Specifically, $\tau_{tel}$(0.88) describes the transmission of the clean optics at the start of the mission while $\tau_{EOL}$(0.8) accounts for the `end of life' transmission where we see losses in the optical train. These transmission coefficients were derived by the manufacturer's cryogenic measurements in an operational environment. $P_{\lambda}$ can be found using: 
\begin{equation}
P_{\lambda} = S_{bgd}(\lambda)\dfrac{A_{tel}}{h\lambda} 
\label{Equation:P}
\end{equation}

\noindent
where $S_{bgd}$ defines the background source flux density. Equation A3 ensures proper unit conversion, providing variables ready to be integrated using equation \ref{Equation:BgdPhotocurrent}.

\begin{table}
  \caption {Specific \textit{JWST} background contributions. Used in conjunction with Equation \ref{Equation:BB}, this information can be used to 
produce a set of profiles describing the wavelength-dependent thermal background emission which will be observed with MIRI.} \label{table:Greybody}
  \begin{center}
    \begin{tabular} { l | l | c }
      \hline\hline
      Component & Emmisivity ($\epsilon$) & Temperature / K  \\
      \hline
      A - Zodiacal& 4.20$\times 10^{-14}$& 5500 \\
      B - Zodiacal & 4.30$\times 10^{-8}$ & \phantom{5}270 \\
      C - DTA Tower & 3.35$\times 10^{-7}$ & \phantom{5}134\\
      D - Primary Mirror & 9.70$\times 10^{-5}$ & \phantom{55}71\\
      E - DTA Shield& 1.72$\times 10^{-3}$ & \phantom{55}62\\
      F - Sun Shield &1.48 $\times 10^{-2}$& \phantom{55}52\\
      G - High Background & 1.31$\times 10^{-4}$ & \phantom{55}87 \\
      
      \hline
    
    \end{tabular}
  \end{center}

\end{table}

Once the background signal is quantified, the noise can be calculated using equation \ref{Equation:Noise}:

\begin{center}
\begin{equation}
N^2_{int} = K_1(i_{sig}+i_{bgd})t_{int} + K_2i_{dark}t_{int} + K_3R^2_N
\label{Equation:Noise}
\end{equation}
\end{center}

\begin{figure}
   \centering
   \includegraphics[width=8cm]{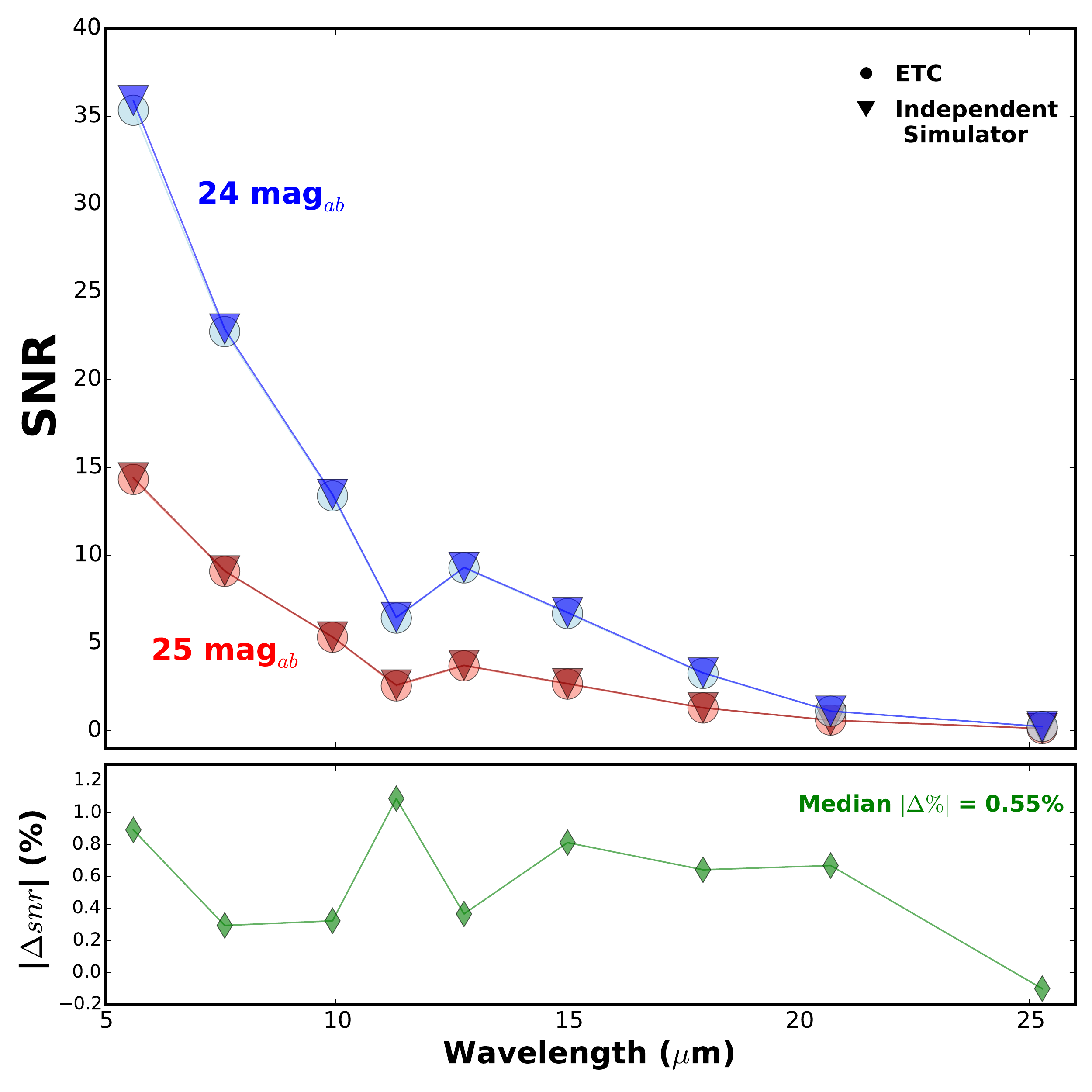}
   \caption{The upper panel presents the average SNR in each MIRI imaging band calculated by both the ETC (filled circles) and our own independent simulator (downward triangles). The red and blue points represent a $m_{AB} \simeq 25$ mag and $m_{AB} \simeq 24$ mag galaxy respectively. The lower panel shows the average SNR percentage offset calculated in each filter (green diamonds), and we also print the median percentage offset for all MIRI bands.\label{fig:sim_test}}
\end{figure} 

\noindent
where it can be seen that the total noise expected to be detected by MIRI consists of three terms, each representing a different background source. The first accounts for the total incident photon flux from the target and background. The second accounts for an induced current detected in each pixel in the absence of illumination (i.e. dark current).
 The third, read noise ($R^2_N$), describes the fluctuations associated with the electronic equipment within the optical systems.

Equation \ref{Equation:K1} defines the factor by which the noise measured within the system exceeds the theoretical shot noise limit:

\begin{center}
\begin{equation}
K_1= K_{exc} \dfrac{6}{5}\dfrac{n^2_{read}+1}{n^2_{read}-1}
\label{Equation:K1}
\end{equation}
\end{center}

\noindent
The shot noise describes the random arrival of photons to the telescope pupil and therefore affects increasingly weaker sources.  Here, it is specifically defined under the conditions that the astronomical target or background is the dominating source of photons. $K_{exc}$ accounts for gain and gain dispersion in the detectors from the signal.  The noise associated with the dark current in term two of equation \ref{Equation:Noise} follows a Poisson form, effectively behaving like shot noise. This allows us to set the coefficient $K_2$ to unity with the dark current ($i_{dark}$) taking a value depending on the observing mode in use (Imager, Low Resolution Spectrometer (LRS), Medium Resolution Spectrometer (MRS), Short Wavelength (SW) or Long Wavelength (LW) Channels) The final term is $n_{read}$ which simply determines the number of equi-spaced pixel reads per integration.

As discussed above, $R_N$ defines the noise picked up from electrical components with a coefficient given by Equation \ref{Equation:K3}:
\begin{center}
\begin{equation}
K_3 = K^2_{RNobs} \dfrac{12n_{read}}{n^2_{read}-1}
\label{Equation:K3}
\end{equation}
\end{center}

\noindent
where $R_N$ is approximately equal to 32.6 electrons or 11.53 electrons depending on whether the instrument is working in `FAST' or `SLOW'  readout mode, respectively. The type of science goal and instrument selected determines which mode is required. `SLOW' mode is mainly used for observing faint objects with MRS spectroscopy, whereas `FAST' mode is sufficient for almost all other observing scenarios. The extra term, $K_{RNobs}$, reflects the factor by which the system noise does not reach the `ideal' behaviour.

Once a target object has been defined in terms of a flux density, these equations enable calculation of  the signal and noise received in each integration. Depending on the instrument (Imager, MRS, or LRS) and the observing mode chosen, specific values for all factors are required. The user can select the readout mode, the number of groups/frames, the number of integrations and the number of exposures for each observation in each photometric band, giving a total exposure time. With the selected background, the simulator delivers the signal-to-noise ratio for the simulated observation of 
the user-defined astronomical target.

Figure\, \ref{fig:sim_test} presents tests completed in order to assess the accuracy of our independent simulator used to calculate and simulate the MIRI observations in this work. In the upper panel, we show the average SNR calculated for two test runs, the first was  a $m_{AB} \simeq 25$ mag object (red points), while the second was a $m_{AB} \simeq 24$ mag object (blue points). The filled downward triangles represent the SNR calculated in a 5550s exposure by our independent simulator, whereas the filled circles represent the SNR calculated using the offical ETC\footnote{\url{https://jwst.etc.stsci.edu}}. The lower panel presents the median SNR percentage offset ($|\Delta snr|$) in each band (green filled diamonds). We assume the same target source, background models and specific observing strategy in all cases. We also appropriately reduce the number of groups/integrations for both the ETC and our independent simulator to avoid background satruation in the longer-wavelength MIRI filters. In general, our independent simulator tends to marginally overestimate the SNR capabilities of MIRI imaging, however this effect is almost negligible producing an average SNR offset of 0.55\%.

\end{document}